# A Set of Virtual Experiments of Fluids, Waves, Thermodynamics, Optics, and Modern Physics for Virtual Teaching of Introductory Physics

*Neel Haldolaarachchige
Department of Physical Science, Bergen Community College, Paramus, NJ 07652

Kalani Hettiarachchilage
Department of Physics, Seton Hall University, South Orange, NJ 07962

January 02, 2020

*Corresponding Author: haldo.physics@gmail.com

**ABSTRACT**

This is the third series of the lab manuals for virtual teaching of introductory physics classes. This covers fluids, waves, thermodynamics, optics, interference, photoelectric effect, atomic spectra and radiation concepts. A few of these labs can be used within Physics I and a few other labs within Physics II depending on the syllabi of Physics I and II classes. Virtual experiments in this lab manual and our previous Physics I (arXiv.2012.09151) and Physics II (arXiv.2012.13278) lab manuals were designed for 2.45 hrs long lab classes (algebra-based and calculus-based). However, all the virtual labs in these three series can be easily simplified to align with conceptual type or short time physics lab classes as desired. All the virtual experiments were based on open education resource (OER) type simulations. Virtual experiments were designed to simulate in-person physical lab experiments. Student learning outcomes (understand, apply, analyze and evaluate) were studied with detailed lab reports per each experiment and end of the semester written exam which was based on experiments. Special emphasis was given to study the student's skill development of computational data analysis.

**CONTENTS**

# EXPERIMENT 1
# FLUIDS: ARCHIMEDES, PASCAL AND BERNOULLI PRINCIPLES

## OBJECTIVE

The Archimedes' principle is studied, and the fluid density is calculated by applying Archimedes principle which also demonstrates the concept of apparent weight of an object. Then the second part of the experiment is demonstrating the Pascal principle. Finally, Bernoulli's principle is investigated.

## THEORY AND PHYSICAL PRINCIPLES

*Archimedes' principle*

When an object of mass M is attached to spring balance is partially or wholly submerged in a liquid the apparent weight, W', that is acting on the string is computed by the product of the apparent mass, M', and gravitational constant, $g$, of 9.81m/s$^2$.

$$W' = M'g \qquad (1)$$

The apparent weight, W', acting on the string due to an object being partially or fully submerged in a liquid can be computed by subtracting the buoyancy force acting on the object, B, from the weight of the object, W, which comprises of the object's mass times the gravitational constant, $g$, of 9.81 m/s$^2$.

$$W' = W - F_B \qquad (2)$$

The buoyancy force, $F_B$, acting on an object in a liquid is computed by the density of the liquid, $\rho_{liquid}$, times the volume of the object submerged, $V_{sub}$, and the gravitational constant, $g$, which is 9.82 m/(s$^2$).

$$F_B = \rho_{liquid}V_{sub}\,g \qquad (3)$$

With the buoyancy force defined, formula 2 can be computed as the following:

$$W' = W - \rho_{liquid}V_{sub}\,g \qquad (4)$$

Hence, density of the fluid can be calculated by using real apparent mass of the object.

$$\rho_{fluid} = \frac{M - M'}{V} \qquad (5)$$

*Pascals' principle*

Pascal principle explains how to use the fluid pressure and displace fluid volume to inside the U-tube for practical application called hydraulic lift.

Consider fluid in a U-tube and there are two objects on pistons (pistons are mass less) as shown in Figure-2.

When two objects are balanced in the same level, pressure just below pistons on both sides must be the same.

$$P_1 = P_2 \rightarrow P_0 + \frac{m_1 g}{A_1} = P_0 + \frac{m_2 g}{A_2} \rightarrow \frac{m_1}{A_1} = \frac{m_2}{A_2} \tag{6}$$

When two pistons are separated by h distance,

$$P_1 = P_2 + h\rho g \tag{7}$$

$$P_0 + \frac{m_1 g}{A_1} = P_0 + \frac{m_2 g}{A_2} + h\rho g \tag{8}$$

$$\frac{m_1}{A_1} = \frac{m_2}{A_2} + h\rho \tag{9}$$

Density of the fluid can be calculated by using Pascals' principle.

$$\rho = \frac{\left(\frac{m_1}{A_1} - \frac{m_2}{A_2}\right)}{h} \tag{10}$$

This also can be done by using a graph of $m_1$ vs h.

$$h = \frac{1}{\rho A_1} m_1 - \frac{m_2}{\rho A_2} \tag{11}$$

Slope of the graph of h vs $m_1$,

$$slope = \frac{1}{\rho A_1} \rightarrow \rho = \frac{1}{A_1 \, slope} \tag{12}$$

*Bernoulli's principle*

Bernoulli's principle explains the work energy principle when fluid is moving from one position to another. At any position of the fluid flow,

$$P + \rho g h + \frac{1}{2}\rho v^2 = constant \tag{13}$$

$$P_1 + \rho g h_1 + \frac{1}{2}\rho v_1^2 = P_2 + \rho g h_2 + \frac{1}{2}\rho v_2^2 \tag{14}$$

When both sides of the tube are in same level,

$$P_1 + \frac{1}{2}\rho v_1^2 = P_2 + \frac{1}{2}\rho v_2^2 \tag{15}$$

## APPARATUS AND PROCEDURE

- Virtual experiment of buoyant force is done by using following:
  http://www.thephysicsaviary.com/Physics/Programs/Labs/ForceBuoyancy/
- A very detail lesson video with detail of data collection with simulator and data analysis with excel can be found here: https://youtu.be/wtXmpdJ-i00

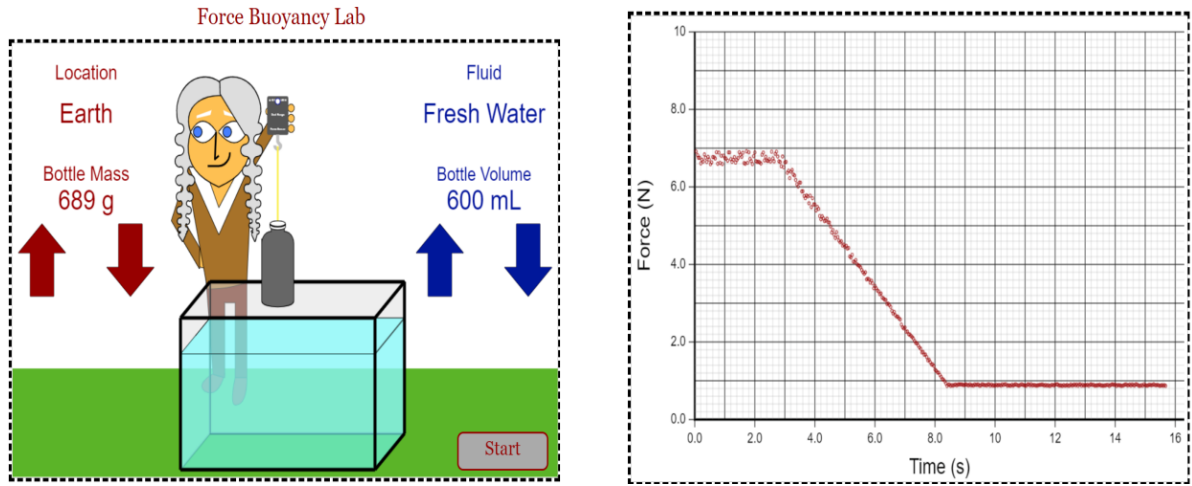

Figure 1 Archimedes simulation (*left*) and force sensor data (*right*)
(Picture credit: http://www.thephysicsaviary.com/)

- Part-A of the experiment is to find the density of fluid by using Archimedes principle.
- Volume of the bottle can be changed by using blue arrows in the right side of the simulation.
- Mass of the bottle can be changed by using red arrows in the left side of the simulation.
- Increase mass of the bottle till the object is fully submerged in the fluid.
- Then find the apparent mass of the object by using the data in the graph of force vs time.
- Find the density of the fluid by using real and apparent mass of the object.
- Repeat above procedure by changing the type of the fluid. To change the fluid type just click on the term "fluid" on the top right side of the simulator.

- Part-B: Virtual experiment of Pascals' principle is done by using following:
- http://www.thephysicsaviary.com/Physics/Programs/Labs/PascalsPrincipleLab/

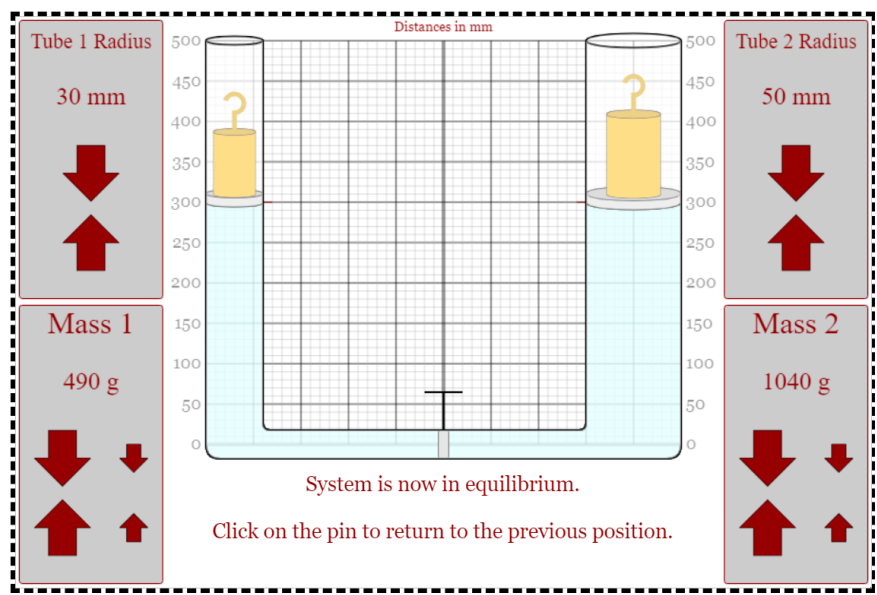

Figure 2 Pascal principle simulation (Picture credit: http://www.thephysicsaviary.com/)

- Red arrows can be used to adjust the values of radius of tubes and values of masses on pistons inside the tubes.
- Set the radius of the tube-1 as 20.0mm and tube-2 as 50.0mm.
- Set the mass $m_2$ on the piston in tube-2 into 1300 gram value.
- Set the mass $m_1$ on the tube-1 into its lowest possible value of 50 gram.
- Release the middle pin and wait until the system goes into equilibrium and then measure the height (h) separation between two pistons.
- Reset the system.
- Then slowly increase the value of mass $m_1$ in tube-1 and measure the separation between the tubes.
- Repeat the procedure for at least 6 different mass-1 values.

- Part-C: Virtual experiment of Bernoulli's equation is done by using following:
- http://www.thephysicsaviary.com/Physics/Programs/Labs/BernoulliLab/
- Adjust the speed of the fluid flow in side-1 to its minimum value.
- Adjust the radius of the side-1 to its maximum value and speed of the flow to minimum value.
- Adjust the radius of the side-2 to its minimum value.
- Measure pressure difference between two sides of the tube.
- Calculate the pressure difference between two sides by using Bernoulli's principle.
- Compare measured and calculated pressure differences by doing percentage differences.
- Then increase speed of the flow (by using up arrow) to maximum value on side-1 and measure the speed of the side-2 and also the pressure on both ends.
- Then decrease the radius of the side-1 (by using the down arrow) and measure the speed of the side-2 and also the pressure on both ends. And repeat this procedure for two more radius values.

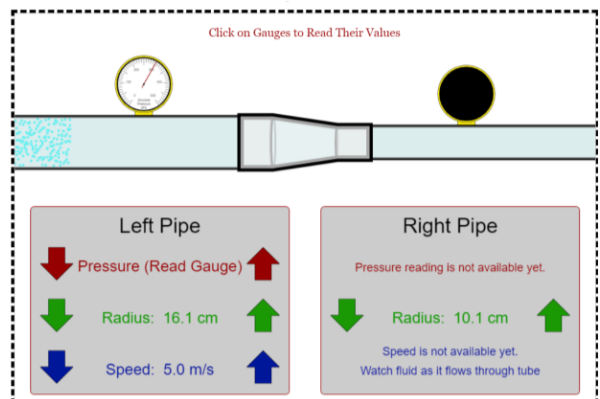

Figure 3 Bernoulli's principle simulation (Picture credit: http://www.thephysicsaviary.com/)

**PRE LAB QUESTIONS**
1) Describe Archimedes principle?
2) Describe buoyant force?
3) Describe apparent mass?
4) Describe Pascals' principle?
5) Describe lamina type fluid follow?
6) Describe continuity of the fluid flow?
7) Describe Bernoulli's principle?

**DATA ANALYSIS AND CALCULATIONS**

*Part A: Calculate density of fluid by using Archimedes principle*

Table 1          Density of fluid calculation

| Name of the fluid and details of the bottle | Force when the object in air [    ] | Force when object is fully submerged in fluid [    ] | Apparent weight of the object [    ] | Buoyant force on the object when fully submerged [    ] | Density of the fluid [    ] |
|---|---|---|---|---|---|
| Fresh Water Mass = Volume= | | | | | |
| Mercury Mass = Volume= | | | | | |
| Gasoline Mass = Volume= | | | | | |
| Maple Syrup Mass = Volume= | | | | | |

Table 2          Percent error of calculated density of fluids

| Name of the fluid [    ] | Calculated density of fluid [    ] | Expected density of fluid [  kg/m$^3$  ] | Percent error analysis [    ] |
|---|---|---|---|
| Fresh Water | | $1.000*10^3$ | |
| Mercury | | $1.360*10^4$ | |
| Gasoline | | $7.800*10^2$ | |
| Maple Syrup | | $1.370*10^3$ | |

*Part B: Investigation of Pascals' principle*

Radius of tube-1 =

Radius of tube-2 =

Mass of tube-2   =

- Increase the mass-1 on piston-1 and measure the height difference between two pistons.
- Calculate the pressure on the piston-1.
- Calculate the and pressure on the other side.
- Compare the pressure on the same level of the fluid.

Table 3          Pascals' principle

| Mass on tube-1 $m_1$ [   ] | Position of piston-1 $h_1$ [   ] | Position of piston-2 $h_2$ [   ] | Separation between pistons, h [   ] |
|---|---|---|---|
|  |  |  |  |
|  |  |  |  |
|  |  |  |  |
|  |  |  |  |
|  |  |  |  |

Table 4          Pressure analysis with Pascal's principle

| Mass on tube-1 $m_1$ [   ] | Pressure on piston-1 $m_1/A_1$ [   ] | Pressure on piston-2 $m_2/A_2$ [   ] | Pressure due to fluid height, h [   ] | Pressure on side-2 [   ] | Percent difference between pressures [   ] |
|---|---|---|---|---|---|
|  |  |  |  |  |  |
|  |  |  |  |  |  |
|  |  |  |  |  |  |
|  |  |  |  |  |  |
|  |  |  |  |  |  |

*Part C: Investigation of Bernoulli's principle.*

Table 5          Bernoulli's principle

| Speed of side-1 [   ] | Radius of the tube of side-1 [   ] | Speed of side-2 [   ] | Radius of the tube of side-2 [   ] | Measured pressure difference [   ] | Calculated pressure difference [   ] | Percent difference [   ] |
|---|---|---|---|---|---|---|
|  |  |  |  |  |  |  |
|  |  |  |  |  |  |  |
|  |  |  |  |  |  |  |
|  |  |  |  |  |  |  |
|  |  |  |  |  |  |  |

# EXPERIMENT 2    STANDING WAVE IN A STRING

## OBJECTIVE

A sine wave generator drives a string vibrator to create a standing wave pattern in a stretched string. The driving frequency and the length, density, and tension of the string are varied to investigate their effect on the speed of the wave in the vibrating string.

## THEORY AND PHYSICAL PRINCIPLES

A stretched string has many natural modes of vibration (three examples are shown below). If the string is fixed at both ends, then there must be a node (place of no amplitude) at each end and at least one antinode (place of maximum amplitude). It may vibrate as a single loop, in which case the length ($L$) of the string is equal to 1/2 the wavelength ($\lambda$) of the wave. It may also vibrate in two loops with a node at each end; then the wavelength is equal to the length of the string. It may also vibrate with a larger integer number of loops. In every case, the length of the string equals some integer number of half wavelengths.

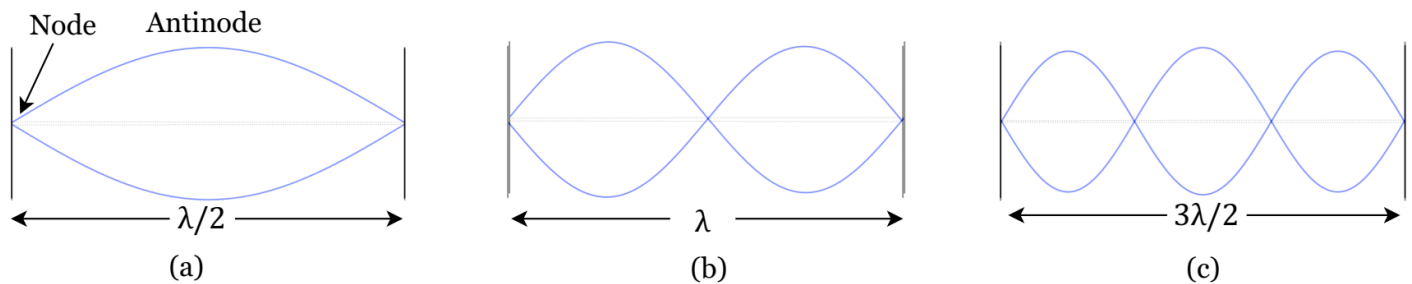

Figure 1   Resonance Modes

If you drive a stretched string at an arbitrary frequency, you will probably not see any particular mode: Many modes will be mixed together. But, if the driving frequency, the tension and the length are adjusted correctly, one vibrational mode will occur at a much greater amplitude than the other modes.

In this experiment, standing waves are set up in a stretched string by the vibrations of an electrically-driven String Vibrator. The arrangement of the apparatus is shown below. The tension in the string ($F_T$) equals the weight of the masses ($mg$) suspended over the pulley. You can alter the tension by changing the mass ($m$).  You can adjust the amplitude ($A$) and frequency ($f$) of the wave by adjusting the output of the Sine Wave Generator, which powers the string vibrator.

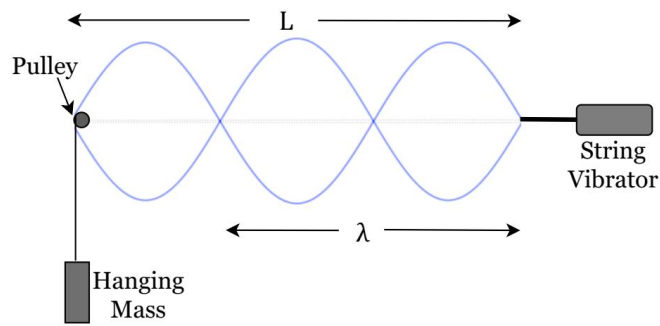

Figure 2   String Vibrator Experimental Setup

$L$ is the length of the vibrating part of the string and λ is the wavelength of the wave.

For the vibrating string with N number of loops,

$$L = N \frac{\lambda}{2} \tag{1}$$

$$\lambda = \frac{2L}{N} \tag{2}$$

For any wave with wavelength λ and frequency $f$, the speed of the wave, v, is

$$v = \lambda f \tag{3}$$

$$v = 2L \frac{f}{N} \tag{4}$$

Wave speed in a string with tension can be written as follows in equation (5). $F_T$ is force of tension ($F_T = mg$, where m is hanging mass, $g$=9.81 m/s$^2$ is an acceleration due to gravity) and linear mass density $\mu = \frac{M}{L}$, M is the mass of the string and L is the length of the string.

$$v = \sqrt{\frac{F_T}{\mu}} \tag{5}$$

By combining equations (4) and (5),

$$f_{cal} = \frac{N}{2L} \sqrt{\frac{F_T}{\mu}} \tag{6}$$

String vibrates at different resonance frequencies with different numbers of loops. Therefore, linear mass density can be found by using the slope of the graph of resonance frequency ($f$) vs number of loops (N).

$$f = \left( \frac{1}{2L} \sqrt{\frac{F_T}{\mu}} \right) N \tag{7}$$

$$\mu = \frac{F_T}{4 L^2 (slope)^2} \tag{8}$$

When resonance happened at fixed frequency but at different tension. Equation (6) can be rewrite as follows.

$$\frac{1}{N} = \frac{1}{2L f \sqrt{\mu}} \sqrt{F_T} \tag{9}$$

Tension force of the string can be found as follows,

$$\sqrt{F_T} = \frac{2L f \sqrt{\mu}}{N} \tag{10}$$

Linear density of the string can be found by using slope of the graph of $\sqrt{F_T}$ vs 1/N.

$$\mu = \left( \frac{slope}{2Lf} \right)^2 \tag{11}$$

**APPARATUS AND PROCEDURE**

- This experiment is done with the following simulation: https://ophysics.com/w8.html
- A very detail lesson video with detail of data collection with simulator and data analysis with excel can be found here: https://youtu.be/f1PLdtdhJ74

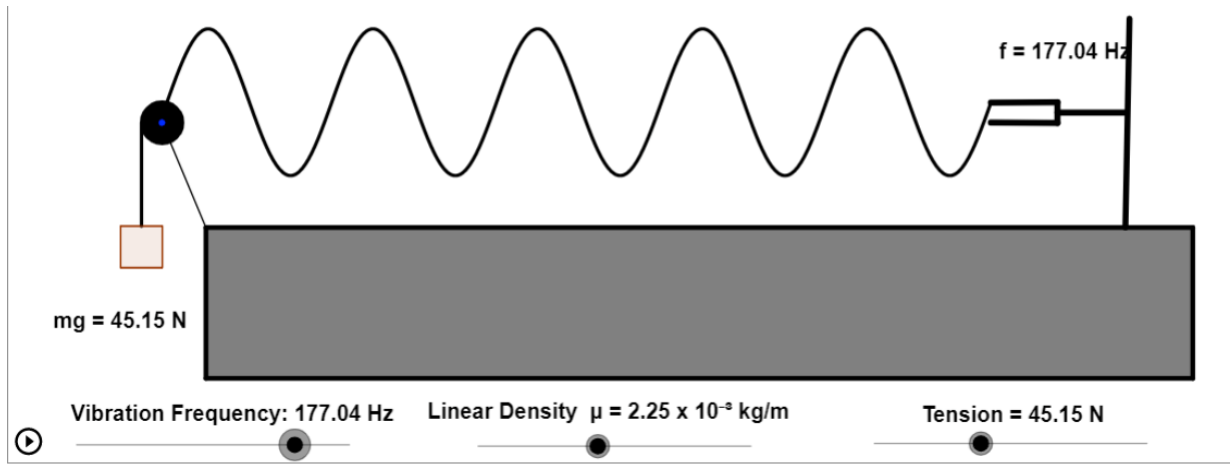

Figure 3  Standing wave on a string simulation (Photo credit: https://ophysics.com/)

*Part A: Effect of frequency*

- Set starting frequency to zero. Set the linear mass density of the string to $\mu = 2.25 \times 10^{-3} \frac{kg}{m}$ and it should be fixed through this part of the experiment.
- Set the tension of the string $F_T$=m$g$=45.15 N and it should be fixed through this part of the experiment.
- Increase frequency ($f$) of the vibrator (by changing vibration frequency of in the simulator) till you observe the smallest complete number of loops in the standing wave.
- To find the exact resonance point you may need to increase frequency ($f$) slowly by observing the amplitude (A) of the wave. Resonance happens when the amplitude reaches maximum. This can be done easily (when close to the resonance point) click on the pause button at the very bottom on the left side then you can see the simulation in slow-motion mode.
- Then note the resonance frequency ($f$) and the number of loops (N).
- Note that you may not get to start with one loop (first resonance mode) instead you may have to start with second (2 loops) or may be third (3 loops) resonance mode.
- Then slowly increase the frequency and find the higher resonance modes and note the resonance frequency and the number of loops in table 1.
- Calculate the frequency of the vibrator ($f_{cal}$) in table-1 by using equation (6) and compare it with applied frequency at resonance.
- Make a graph of calculated frequency ($f_{cal}$) vs number of loops (N) on the string.
- Do the linear fitting and calculate linear density of string $\mu_{graph}$ by using the slope of the graph.
- Find the PD between $\mu_{graph}$ and $\mu = 2.25 \times 10^{-3} \frac{kg}{m}$?

*Part B: Effect of tension of the string*

- Set the starting frequency to *f*=75.15 Hz and it be fixed during this part of the experiment.
- Set the linear density of the string to $\mu = 2.25 \times 10^{-3} \frac{kg}{m}$ and it should be fixed through this part of the experiment.
- Set the starting tension of the string to the lowest possible value in the simulator.
- Increase the tension of the string slowly till the resonance point. Note the number of loops (N) and tension ($F_T$) in table 2. Repeat with increasing tension and for each resonance point note N and $F_T$.
- Make a graph of square-root of tension force $\sqrt{F_T}$ vs 1/N.
- Do the linear fitting and calculate linear density of string $\mu_{graph}$ by using the slope of the graph (see equation-11).
- Find the PD between $\mu_{graph}$ and $\mu = 2.25 \times 10^{-3} \frac{kg}{m}$ ?

*Part C: Effect on resonance frequency at lower and higher linear densities of string*

- Set starting frequency to zero. Set the linear mass density of the string to $\mu = 1.25 \times 10^{-3} \frac{kg}{m}$ and it should be fixed through this part of the experiment.
- Set the tension of the string $F_T$=m*g*=45.15 N and it should be fixed through this part of the experiment.
- Increase frequency (*f*) of the vibrator till you observe a certain number of complete loops in the standing wave.
- The note the frequency (*f*) of vibrator at resonance and the number of loops (N) in table 3
- Make a graph of frequency of vibrator at resonance (*f*) vs number of loops (N) on the string.
- Do the linear fitting and calculate linear density of string $\mu_{graph}$ by using the slope of the graph.
- Find the PD between $\mu_{graph}$ and $\mu = 1.25 \times 10^{-3} \frac{kg}{m}$?
- Repeat all the above for higher linear density $\mu = 3.55 \times 10^{-3} \frac{kg}{m}$ of string and note the data in table 4.

## PRE LAB QUESTIONS

1) How is wave speed related to frequency and wavelength?
2) How is the period of oscillation related to wave speed?
3) What is a standing wave?
4) Explain nodes and antinodes of a standing wave?
5) What are normal modes or resonance states?
6) How does the wavelength of a standing wave in a vibrating string vary with the tension force in the string and/or the linear mass density of the string?
7) How many normal modes of oscillation or natural frequencies does each of the following have:
   - A simple pendulum.
   - A spring oscillator

# DATA ANALYSIS AND CALCULATIONS

*Part A: Effect of frequency*

- Length of the string between vibrator and pulley =

- Applied tension force $F_T$ =

- Linear mass density of string $\mu$ =

<div align="center">Table 1          Frequency and number of loops at resonance</div>

| Frequency of vibrator at resonance $f$ [     ] | Number of loops $N$ | Calculated frequency of vibrator $f_{cal} = \left( \dfrac{1}{2L} \sqrt{\dfrac{F_T}{\mu}} \right) N$ [     ] | PD Between $f$ and $f_{cal}$ [     ] |
|---|---|---|---|
|  |  |  |  |
|  |  |  |  |
|  |  |  |  |
|  |  |  |  |
|  |  |  |  |
|  |  |  |  |

- Make a graph of calculated frequency ($f_{cal}$) vs number of loops (N) on the string.
- Do the linear fitting and calculate linear density of string $\mu_{graph}$ by using the slope of the graph (equation 8).
- Find the PD between $\mu_{graph}$ and $\mu = 2.25 \times 10^{-3} \frac{kg}{m}$?

*Part B: Effect of tension*

- Length of the string between vibrator and pulley =

- Applied frequency of the vibrator $f$ =

- Linear mass density of string $\mu$ =

Table 2     Tension force and number of loops at resonance

| Number of loops at resonance N | Tension force at resonance $F_T$ [    ] | 1/N | $\sqrt{F_T}$ [    ] |
|---|---|---|---|
|  |  |  |  |
|  |  |  |  |
|  |  |  |  |
|  |  |  |  |
|  |  |  |  |
|  |  |  |  |

- Make a graph of square-root of tension force $\sqrt{F_T}$ vs 1/N.
- Do the linear fitting and calculate linear density of string $\mu_{graph}$ by using the slope of the graph (equation 11).
- Find the PD between $\mu_{graph2}$ and $\mu = 2.25 \times 10^{-3} \frac{kg}{m}$ ?

*Part C: Effect of frequency at lower linear density of string*

- Length of the string between vibrator and pulley =
- Applied tension force $F_T$ =
- Linear density of string $\mu$ =

Table 3   Frequency and number of loops at resonance for lower linear density of string

| Frequency of vibrator at resonance $f$ [     ] | Number of loops $N$ |
|---|---|
|  |  |
|  |  |
|  |  |
|  |  |
|  |  |
|  |  |

- Make a graph of frequency ($f$)  vs number of loops (N) on the string.
- Do the linear fitting and calculate linear density of string $\mu_{graph}$ by using the slope of the graph (equation 8).
- Find the PD between $\mu_{graph}$ and $\mu = 1.25 \times 10^{-3} \frac{kg}{m}$?

- Linear density of string μ =

Table 4   Frequency and number of loops at resonance for higher linear density of string

| Frequency of vibrator at resonance $f$ [     ] | Number of loops $N$ |
|---|---|
|  |  |
|  |  |
|  |  |
|  |  |
|  |  |
|  |  |

- Make a graph of frequency ($f$)  vs number of loops (N) on the string.
- Do the linear fitting and calculate linear density of string $\mu_{graph}$ by using the slope of the graph (equation 8).
- Find the PD between $\mu_{graph}$ and $\mu = 3.55 \times 10^{-3} \frac{kg}{m}$?

# EXPERIMENT 3    SOUND WAVES AND RESONANCE AIR COLUMN

## OBJECTIVE

Longitudinal waves are investigated. Sound waves and resonance modes with close end tube are used. Sound speed is analyzed with temperature and with different gas mediums.

## THEORY AND PHYSICAL PRINCIPLES

*Resonance of close tube*

Air column inside one end closed tube resonates in fundamental resonance mode as shown in Figure 1. Air molecules at the close end do not vibrate therefore it looks like in the node on standing wave. Air molecules at the open end vibrate at maximum amplitude making antinodes like in the standing wave.

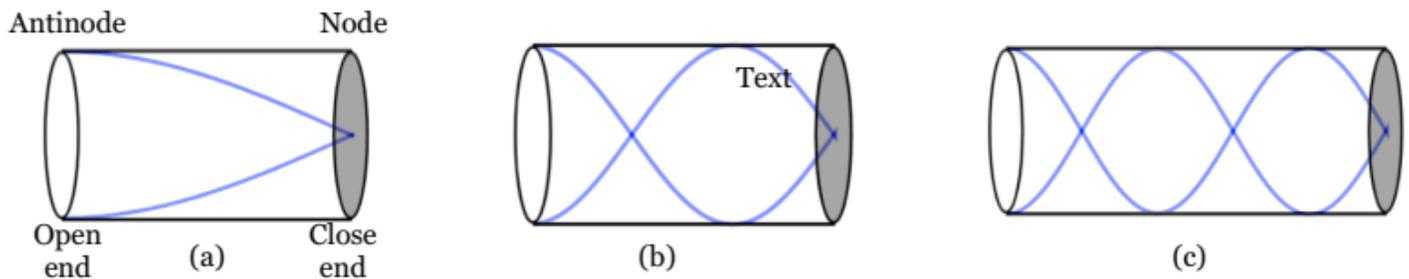

Figure 1   Standing wave modes with one end closed tube

Air column inside the close tube can create other (or higher order) resonance modes, which means there can be a resonance as far as the open end is antinode and the close end is a node. This can happen any odd integer multiplication of quarter wavelength.

$$\lambda = \frac{4}{n}L, \quad n = 1, 3, 5, 7,\ldots\ldots\ldots \tag{1}$$

L is the length of the tube.

Speed of the sound can be calculated, $v = \lambda f = \frac{4}{n}Lf_n$ $\tag{2}$

$$f_n = \frac{nv}{4}L \quad , n = 1, 3, 5, 7,\ldots\ldots\ldots \tag{3}$$

By using variable frequency and measuring the tube length for fundamental resonance, it is possible to find the speed of the sound by using a graph of frequency vs length.

*Resonance of open end tube*

Air column inside the open tube can create other (or higher order) resonance modes, which means there can be a resonance as far as open ends have antinode. This can happen any integer multiplication of half wavelength.

$$\lambda = \frac{2}{n}L, \quad n = 1, 2, 3, 4,,\ldots\ldots\ldots \tag{4}$$

L is the length of the tube.

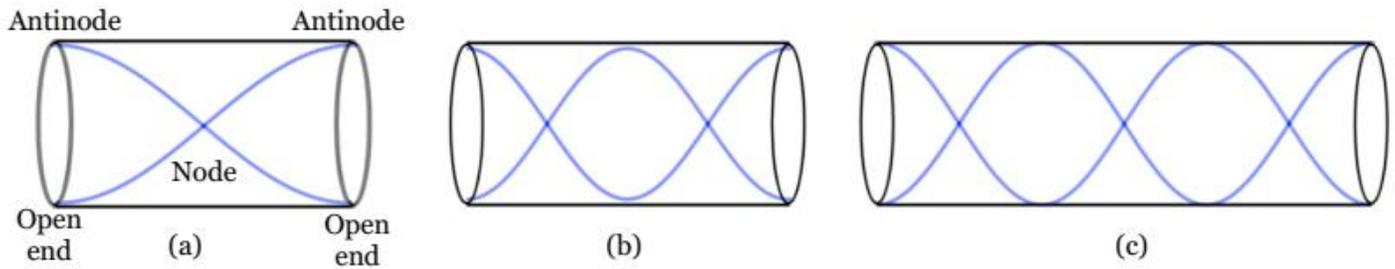

Figure 2   Standing wave modes with both ends open tube

Speed of the sound can be calculated, $v = \lambda f = \frac{2L}{n} f_n$ (5)

$$f_n = \frac{nv}{2} L \quad , \text{n = 1, 2, 3, 4,,}\ldots\ldots\ldots$$ (6)

By using variable frequency and measuring the tube length for fundamental resonance, it is possible to find the speed of the sound by using a graph of frequency vs length.

**APPARATUS AND PROCEDURE**

- This experiment is done with following simulation:
  http://amrita.olabs.edu.in/?sub=1&brch=5&sim=36&cnt=4
- A very detail lesson video with detail of data collection with simulator and data analysis with excel can be found here: https://youtu.be/iZcc0G_VVcw

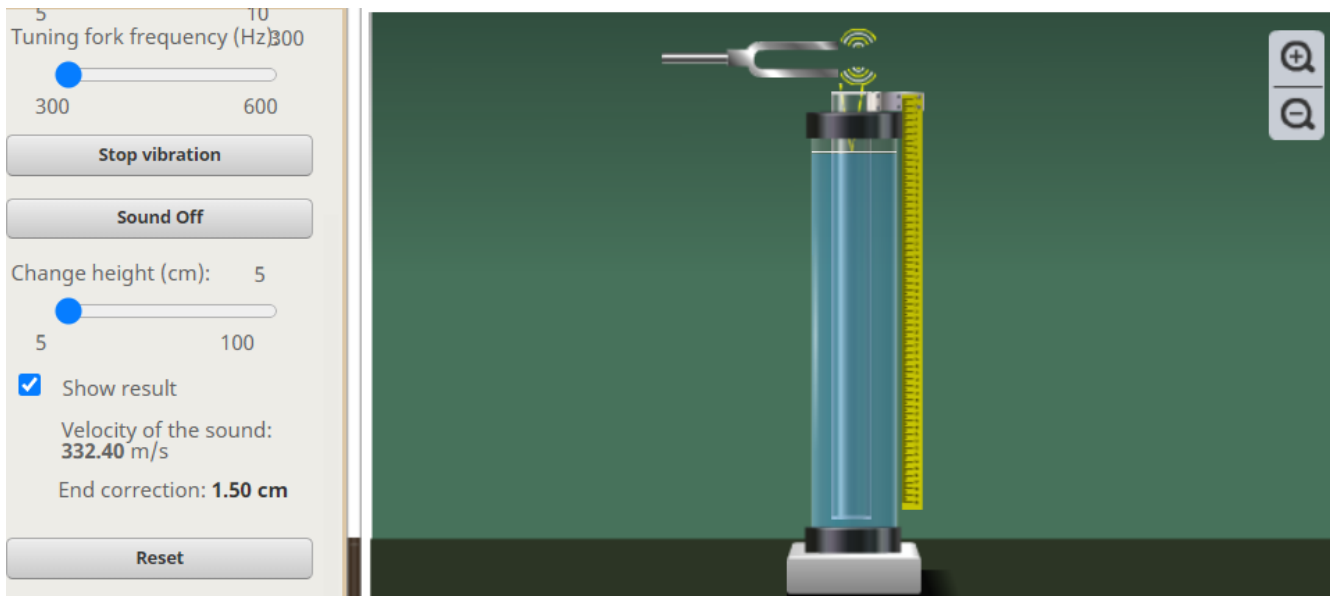

Figure 3    Close end tube resonance simulation
(Picture credit: http://amrita.olabs.edu.in/)

*Part A: Speed of the sound with close tube first resonance*

- Set surrounding to "air".
- Set surrounding temperature to standard room temperature of 25 $^0$C.
- Set tube diameter and tuning fork frequency to lowest values.
- Click hit tuning fork and sound on.
- Slowly increase the length of the innertube till you hear resonance (a beep).
- Note the length frequency and the length of the tube at resonance in table-1.

*Part B: Speed of the sound with close tube second resonance*

- Set surrounding to "air".
- Set surrounding temperature to standard room temperature of 25 $^0$C.
- Set tube diameter and tuning fork frequency to lowest values.
- Click hit tuning fork and sound on.
- Slowly increase the length of the innertube till you hear a second resonance (second beep sound).
- Note the length frequency and the length of the tube at resonance in table-2.

*Part C: Temperature effect of sound speed*

- Set surrounding to "air".
- Set surrounding temperature to its lowest value.
- Set tube diameter and tuning fork frequency to lowest values.
- Click hit tuning fork and sound on.
- Slowly increase the length of the innertube till you hear the first resonance (a beep sound).
- Then increase the temperature values given in table-3 and measure the length of the tube at first resonance.

**PRE LAB QUESTIONS**

1) Describe the sound motion in terms of pressure wave?
2) Describe the resonance modes of open-end tube?
3) Describe the resonance modes in close-end tube?
4) Does the sound speed depend on temperature?
5) Suppose that in this experiment the temperature of the room had been lower; what effect would this have had on the distance between nodes for each reading?  Explain.

**DATA ANALYSIS AND CALCULATIONS**

*Part A: Speed of the sound with close tube first resonance*

Table 1     Frequency vs wavelength to find the speed of the sound for first resonance

| Frequency $f$ [    ] | Length of the tube $L$ [    ] | Wave Length $\lambda$ [    ] | $1/\lambda$ [    ] | Speed of sound $v_{cal}$ [    ] |
|---|---|---|---|---|
|  |  |  |  |  |
|  |  |  |  |  |
|  |  |  |  |  |
|  |  |  |  |  |
|  |  |  |  |  |
|  |  |  |  |  |

- Calculate the speed of the sound for each of the wavelengths and frequencies.
- Calculate the average speed of the sound by averaging the last column $(v_{avg})$?
- Make a graph of frequency (*f*) vs 1/$\lambda$?
- Fit the data with linear fitting.
- Calculate the speed of the sound $(v_{graph})$ by using the slope of the above graph?
- Calculate the standard value of speed of the air $\left[ v_{air} = 331\frac{m}{s} + \left( 0.6 \times \frac{T}{C^0} \right) \right]$?
- Calculate the percent error of $(v_{avg})$ vs $(v_{air})$?
- Calculate the percent error of $(v_{graph})$ vs $(v_{air})$?

*Part B: Speed of the sound with close tube second resonance*

- Calculate the speed of the sound for each of the wavelengths and frequencies.
- Calculate the average speed of the sound by averaging the last column $(v_{avg})$?
- Make a graph of frequency (*f*) vs 1/$\lambda$?
- Fit the data with linear fitting.
- Calculate the speed of the sound $(v_{graph})$ by using the slope of the above graph?
- Calculate the standard value of speed of the air $\left[ v_{air} = 331\frac{m}{s} + \left( 0.6 \times \frac{T}{C^0} \right) \right]$?
- Calculate the percent error of $(v_{avg})$ vs $(v_{air})$?
- Calculate the percent error of $(v_{graph})$ vs $(v_{air})$?

Table 2      Frequency vs wavelength to find the speed of the sound for second resonance

| Frequency $f$ [    ] | Length of the tube $L$ [    ] | Wavelength $\lambda$ [    ] | $1/\lambda$ [    ] | Speed of sound $v_{cal}$ [    ] |
|---|---|---|---|---|
|  |  |  |  |  |
|  |  |  |  |  |
|  |  |  |  |  |
|  |  |  |  |  |
|  |  |  |  |  |
|  |  |  |  |  |

*Part C: Temperature effect of sound speed*

Table 3      Sound speed at difference temperature values

| Temperature [   $^0$C   ] | Length of the tube $L$ [    ] | Wavelength $\lambda$ [    ] | Speed of the sound $v_{cal}$ |
|---|---|---|---|
| 2.00 |  |  |  |
| 25.0 |  |  |  |
| 50.0 |  |  |  |
| 75.0 |  |  |  |
| 95.0 |  |  |  |

- Make a graph of speed of sound (*v*) vs temperature (T).
- Discuss the behavior of the graph.

*Part D: Medium effect of sound speed*

Table 4      Sound speed with different gas in the tube

| Gas type [    ] | Length of the tube $L$ [    ] | Wavelength $\lambda$ [    ] | Speed of the sound $v_{cal}$ |
|---|---|---|---|
|  |  |  |  |
|  |  |  |  |
|  |  |  |  |
|  |  |  |  |
|  |  |  |  |

- Discuss the effect of the gas type into the speed of the sound.

# EXPERIMENT 4    HEAT CAPACITY AND LATENT HEAT

## OBJECTIVE

Specific heat capacity of unknown metals and liquids is investigated by using a calorimetry method. Latent heat of fusion is calculated.

## THEORY AND PHYSICAL PRINCIPLES

*Calorimetry and zeroth law of the thermodynamics*

When a hot object at high temperature connects with a cold object at low temperature thermal energy exchanges between the objects. This is called the zeroth law of thermodynamics, which states that when objects at different temperatures are thermally connected then heat energy transfers from hot object to cold object till the system reaches thermal equilibrium. If the system is isolated from the environment, then there is not any energy loss and total energy loss from hot objects should be absorbed by the cold object. Energy conservation can be applied to heat exchange between hot and cold objects.

*Heat capacity*

When heat energy (Q) is absorbed or lose by an object its temperature changes ($\Delta T$). It is observed that the temperature change of the object is directly proportional to the amount of heat exchange. Proportionality constant is called the heat capacity (C) of the object and the heat capacity is directly proportional to mass (m) of the object. Heat capacity per unit mass is called the specific heat capacity (c) of the materials.

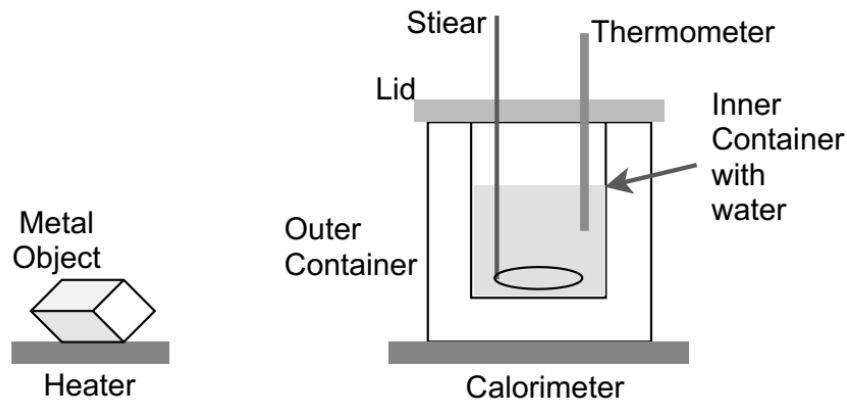

Figure 1   Calorimetry experiment. Heated object is transferred into a beaker with water. Final temperature is measured at the thermal equilibrium of the system

$$Q \propto \Delta T \rightarrow Q = C\Delta T \qquad (1)$$

$$Q = cm\Delta T \qquad (2)$$

Units of thermal specific heat capacity is $\frac{J}{kg\ ^\circ C}$.

$$Q(water) + Q(container) + Q(metal) = 0 \tag{3}$$

$$c_w m_w (T_f - T_i) + c_c m_c (T_f - T_i) + c_o m_o (T_f - T_H) = 0 \tag{4}$$

$c_w$ is heat capacity of water, $c_c$ is heat capacity of calorimeter inner container, $c_o$ is heat capacity of metal object, $m_w$ is mass of water, $m_c$ is mass of calorimeter inner container, $m_0$ is mass of metal object, $T_i$ is initial temperature of water and calorimeter, $T_f$ is final or equilibrium temperature of system (water + calorimeter + metal object), $T_H$ is initial temperature of metal object.

By solving the above equation a specific heat capacity of the metal object can be found.

$$c_o = \frac{c_w m_w (T_f - T_i) + c_c m_c (T_f - T_i)}{m_o (T_H - T_f)} \tag{5}$$

It is important to point out that in this experiment, heat lost from a hot object is fully absorbed by liquid in the calorimeter. There is no calorimeter material to consider in the simulator.

$$c_o = \frac{c_w m_w (T_f - T_i)}{m_o (T_H - T_f)} \tag{6}$$

*Latent heat*

Material can absorb or release thermal energy without changing its temperature when it undergoes a thermal phase transition such as solid to liquid (or liquid to solid) and liquid to vapor (or vapor to liquid).

An amount of heat absorbed or lost during phase transition is directly proportional to amount of mass.

$$Q \propto m \rightarrow Q = L\,m \tag{7}$$

$$L = \frac{Q}{m} \tag{8}$$

Units of latent heat is $\frac{J}{kg}$.

Latent heat is the amount of heat needed to change the phase of one kilogram of material. Latent heat can be two types as follows.

- Latent heat of fusion = $L_f$ = solid changes to liquid (or liquid changes to solid)
- Latent heat of vaporization = $L_v$ = liquid changes to vapor (or vapor changes to liquid)

Consider an ice piece at zero-degree temperature drop into a calorimeter with water.

$$Q_1(water) + Q_2(container) + Q_3(ice\ to\ water) + Q_4(ice - water) = 0 \tag{9}$$

$Q_1$(water) = heat absorbs by the water in the calorimeter

$Q_2$(container) = heat absorbs by the water in the calorimeter material

$Q_3$(ice to water) = heat needed to change zero-degree ice to zero-degree water, (phase change)

$Q_4$(ice-water) = zero-degree water (from ice) changes to final temperature of the system

$$c_w m_w (T_f - T_i) + c_c m_c (T_f - T_i) + L_f m_i + c_w m_i (T_f - 0) = 0 \qquad (10)$$

By solving the above equation latent heat can be found.

$$L_f = \frac{-c_w m_w (T_f - T_i) - c_c m_c (T_f - T_i) - c_w m_i T_f}{m_i} \qquad (11)$$

It is important to point that in this experiment, there is no calorimeter material to consider in the simulator. Heat exchange is between water in the calorimeter and ice.

$$L_f = \frac{-c_w m_w (T_f - T_i) - c_w m_i T_f}{m_i} \qquad (12)$$

## APPARATUS AND PROCEDURE

*Part A: Specific heat*

- This experiment is done with the following simulation:
  https://media.pearsoncmg.com/bc/bc_0media_chem/chem_sim/calorimetry/Calor.php
- A very detail lesson video with detail of data collection with simulator and data analysis with excel can be found here: https://youtube.com/playlist?list=PLsVLYnCPRO5kT2jveoDCLcaUnznuihNqc

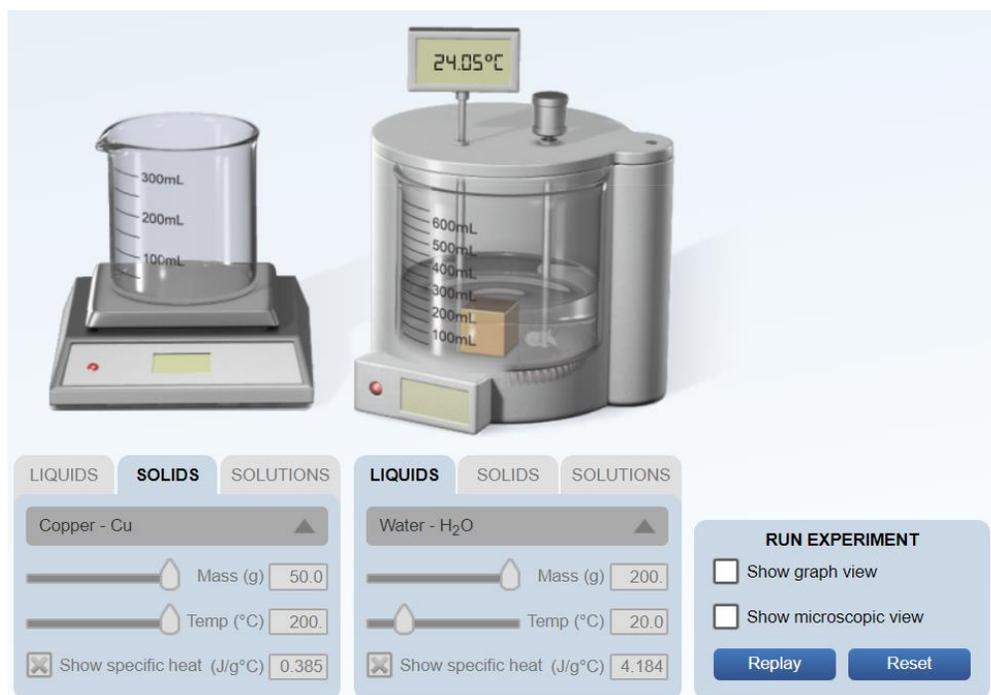

Figure 2    Simulation of thermal expansion

- Select the "experiment" on the top selection menu.
- Click on "solid" on the selection panel on the left side bottom corner.
- Select "unknown metal I" on selection tool.
- Select the to 50.0g and select the temperature to 200.0 ⁰C.
- Click the "next" button on the very left side bottom corner on the simulator.
- Select "liquid" on the middle section panel (below the calorimeter).
- Select the "water" and mass to 200.0 g. Temperature of the water is auto set to 20.0 ⁰C.

- Click on "show graph view" and then click the start button on the very right-side bottom corner.
- Hot object is dropped to a calorimeter and the graph shows the temperature variation of two objects as a function of time.
- When the system reaches equilibrium, the calorimeter scale shows equilibrium temperature.
- Reset the simulator and repeat the procedure for the second unknown metal object.
- Reset the simulator and repeat above procedure for unknown liquid I and II.

*Part B: Latent heat of fusion*

- Reset the simulator
- Click on "solid" on the selection panel on the left side bottom corner.
- Select "ice" on the selection tool which is at the very last of the selection.
- Select the to 30.0g and select the temperature to 0 $^0$C.
- Click the "next" button on the very left side bottom corner on the simulator.
- Select "liquid" on the middle section panel (below the calorimeter).
- Select the "water" and mass to maximum possible value. Temperature of the water set to 20.0 $^0$C.
- Click on "show graph view" and then click the start button on the very right-side bottom corner.
- Ice is dropped to a calorimeter and the graph shows how the temperature variation of two objects as a function of time.
- When the system reaches equilibrium temperature, calorimeter scale shows equilibrium temperature.
- Repeat the above procedure by selecting 40.0g of ice.

PRE LAB QUESTIONS
1) Describe the heat capacity and specific heat capacity?
2) Describe the latent heat of fusion?
3) Describe the latent heat of vaporization?
4) Describe what happens to the heat absorb/loss during the phase change? (what really happening in the system at microscopic level because there is no temperature change during phase change)

**DATA ANALYSIS AND CALCULATIONS**

*Part A: Thermal Expansion*

Table 1      Calorimetric measurements of unknown metal and unknown liquid

| | Mass of the unknown object/liquid $m_0$ [    ] | Initial temperature of hot object/liquid $T_H$ [    ] | Information of calorimeter liquid (water) | |
| --- | --- | --- | --- | --- |
| | | | Mass of the water $m_w$ [    ] | Initial temperature of water $T_i$ [    ] |
| Unknown metal I $c_0$(given)= $c$(water)= | | | | |
| Unknown metal II $c_0$(given)= $c$(water)= | | | | |
| Unknown liquid I $c_0$(given)= $c$(Acetone)= | | | | |
| Unknown liquid II $c_0$(given)= $c$(Acetone)= | | | | |

Table 2      Heat capacity analysis of unknown metal/liquid by calorimetry

| | Specific heat capacity expected (given from simulator) $c_0$ (given) [    ] | Specific heat capacity calculated $c_0$ (cal) [    ] | PE between $c_0$ (given) and $c_0$ (cal) [    ] |
| --- | --- | --- | --- |
| Unknown metal I | | | |
| Unknown metal II | | | |
| Unknown liquid I | | | |
| Unknown liquid II | | | |

Table 3    Heat exchange analysis of unknown metal and unknown liquid

|  | Heat loss from hot object or liquid [    ] | Heat absorb by the water in the calorimeter [    ] | PE between heat loss and heat absorb [    ] |
|---|---|---|---|
| Unknown metal I |  |  |  |
| Unknown metal II |  |  |  |
| Unknown liquid I |  |  |  |
| Unknown liquid II |  |  |  |

*Part B: Latent heat*

Table 4    Analysis of latent heat of fusion

| Mass of the ice, $m_i$ [    ] | Initial temperature of ice, $T_i$(ice) [    ] | Mass of the water in calorimeter, $m_w$ [    ] | Initial temperature of calorimeter, $T_i$ [    ] | Final temperature of the system, $T_f$ [    ] |
|---|---|---|---|---|
|  |  |  |  |  |
|  |  |  |  |  |
| Latent heat calculated $L_1$, [    ] | Latent heat calculated $L_2$, [    ] | Latent heat calculated average $L_{avg}$, [    ] | Latent heat known $L_{known}$ [    ] | PE between known and calculated average |
|  |  |  |  |  |
| Total amount of heat absorbed by the amount of ice to reach final temperature [    ] | | Total amount heat loss by the water in the calorimeter [    ] | | PD between heat absorb and heat loss |
|  | | | | |

# EXPERIMENT 5   GAS LAWS AND KINETIC THEORY OF GAS

## OBJECTIVE

Gas laws are investigated. Boyle's law is investigated by measuring gas pressure as a function of volume at constant temperature. Charles's law is investigated by measuring volume as a function of temperature at constant pressure. Average speed of gas molecules is investigated as a function of temperature.

## THEORY AND PHYSICAL PRINCIPLES

Robert Boyle (1627 - 1691) investigated gas volume at constant temperature with respect to pressure and found inverse proportionality of them.

$$V \propto \frac{1}{P} \tag{1}$$

Jacques Charles (1746 – 1823) investigated gas volume at constant pressure with respect to temperature and found linear behavior.

$$V \propto T \tag{2}$$

By combining Boyle and Charles laws,

$$\frac{PV}{T} = constant \tag{3}$$

This generally refer as ideal gas law, $\frac{PV}{T} = Nk_B$ $\tag{4}$

N is number of molecules and $k_B = 1.38 \times 10^{-23} \frac{J}{K}$

$$N = N_A \, n \tag{5}$$

N is number of atoms or molecules, $n$ is number of mole and N$_A$ is Avogadro number, $N_A = 6.02 \times 10^{23} \frac{1}{mol}$

Ideal gas law generally written with universal gas constant, R.

$$R = N_A \, k_B \; = \; 8.31 \frac{J}{mol \, K} \tag{6}$$

$$PV \; = \; nRT \tag{7}$$

Ideal gas law was modified by Van der Waals and he introduced an equation of state, which can be applied for real gas.

$$\left[ P + a \left( \frac{n}{V} \right)^2 \right] (V - nb) = nRT \tag{8}$$

$a$ and $b$ are constants depending on the type of gas and should be determined experimentally for each gas.

Pressure can be explained by using microscopic views of motion of gas molecules.

$$P = \frac{1}{3} \frac{Nm\bar{u}^2}{V} \tag{9}$$

$m$ is mass of gas molecule and $\bar{u}$ is the average speed of a gas molecule.

$$PV = Nk_BT = \frac{1}{3}Nm\bar{u}^2 \tag{10}$$

Average kinetic energy of gas molecule,

$$KE = \frac{3}{2}k_BT \tag{11}$$

Internal energy of gas, $E = N(KE) = \frac{3}{2}Nk_BT$ $\tag{12}$

$$E = \frac{3}{2}nRT \tag{13}$$

Root-mean-square (rms) speed of gas molecule, $u_{rms}$

$$u_{rms} = \sqrt{\bar{u}^2} = \sqrt{\frac{3k_BT}{M}} \tag{14}$$

M is the molar mass of gas.

## APPARATUS AND PROCEDURE

- This experiment is done with following simulation:
  https://phet.colorado.edu/sims/html/gas-properties/latest/gas-properties_en.html
- A very detail lesson video with detail of data collection with simulator and data analysis with excel
  can be found here: https://youtu.be/5G4Rxyuc2Ec

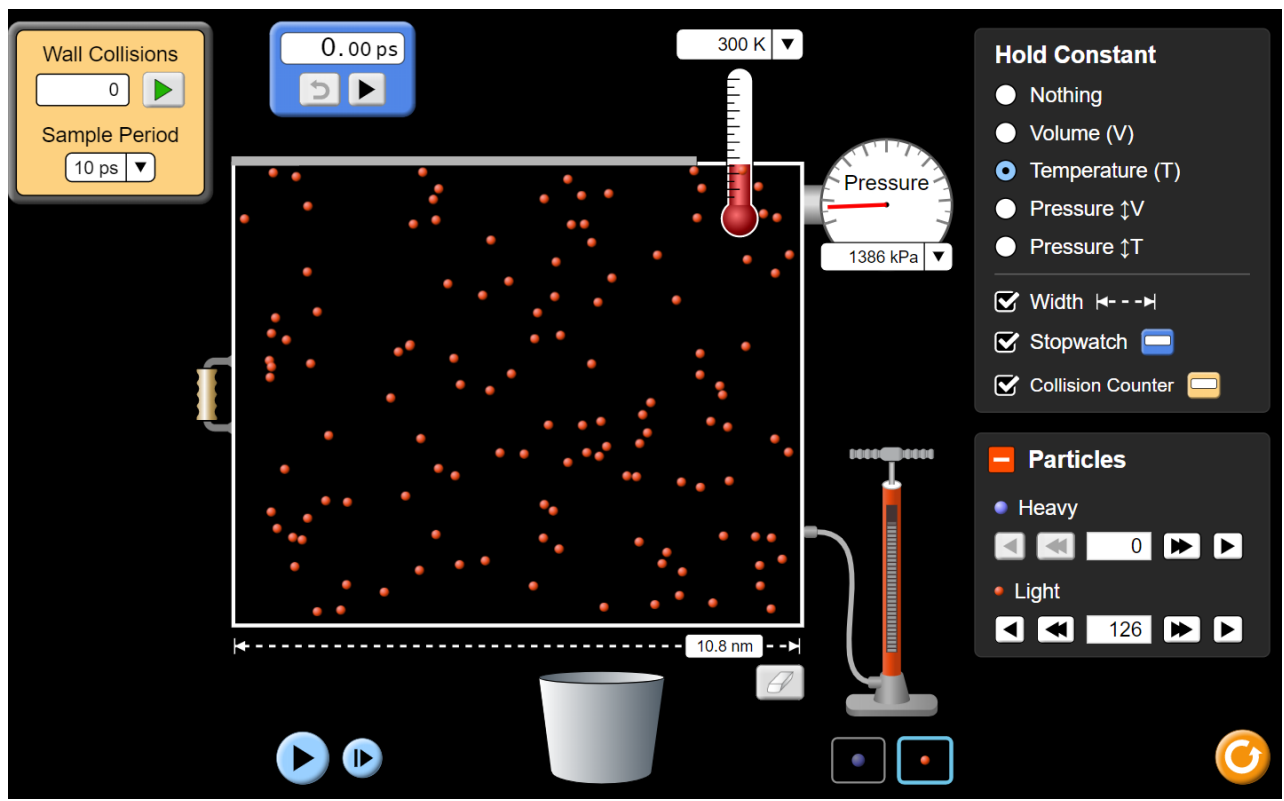

Figure 1     Simulation for analysis of gas laws (Picture credit: https://phet.colorado.edu/)

*Part A: Study Boyle's law behavior of gas*

- Gas pressure is measured as a function of changing volume of the container by using option "Ideal" in the simulator.
- Click on the handle shape icon attached to the middle left side of the container and drag it to the largest volume of the container.
- Select the particle type by clicking the orange color at the bottom of the gas pump.
- Click on the pump icon to add gas into the container. Add gas till the pressure is about 12 atm. You can use the unit conversion at the top of the column as 1.0atm = $1.01*10^5$ Pa.
- Click on the "pause" icon in the bottom left.
- Click on temperature icon in "hold constant" to top right corner
- Note down volume and pressure. Volume should be calculated W*L*T. W=15.0nm, L=10.0nm and T-5.0 nm.
- Then change the volume of the container each 2.0nm in width and note the pressure. Consider that the length and thickness of that container remains unchanged.
- Make a graph of volume vs pressure and discuss the behavior of the graph in terms of gas laws.
- Make a graph of volume vs 1/pressure and discuss the behavior of the graph in terms of gas laws.

*Part B: Study Charles law behavior of gas*

- Gas temperature is measured as a function of changing volume of the container.
- Use the same orange particle type and same amount of gas added in the previous section.
- Click on the pressure (with T) icon in "hold constant" to the top right corner.
- Use the amount of gas added in previous case (A).
- Click on the handle shape icon attached to the middle left side of the container and drag it to the largest volume of the container.
- Click on the "pause" icon on the bottom left.
- Note down volume and temperature. Volume should be calculated W*L*T. W=5.0nm, L=10.0nm and T-5.0 nm.
- Then change the volume of the container each 2.0nm in width and note the temperature. Consider that the length and thickness of that container remains unchanged.
- Make a graph of volume vs temperature and discuss the behavior of the graph in terms of gas laws.

*Part C: Study temperature vs pressure behavior of gas*

- Gas pressure is measured as a function of changing the temperature of the container.
- Use the same orange particle type and same amount of gas added in the previous section.
- Click on the volume icon in "hold constant" to the top right corner.
- Use the amount of gas added in previous case (A).
- Leave the container width at the smallest value of 5.0nm.
- And click on the "Volume" in the top right corner to hold the volume constant.
- Click on the "play" icon in the bottom left.
- Then change the temperature to about 100K of the gas chamber by using the container icon below the gas chamber. Click and drag the slider to the blue color side to cool the gas container.
- Click the pause button and note down the pressure and temperature.
- Click the play button and then change the temperature of the gas chamber by 50K (using heating icon).

- Click the pause button and note down the pressure and play button. And repeat the last two steps by increasing temperature 50K at a time.
- Make a graph of pressure vs temperature and discuss the behavior of the graph in terms of gas laws.

*Part D: Study average speed of gas molecule as a function of temperature and pressure*

- This part of the experiment should be done by selecting the "Energy" of the selection panel at the very bottom of the simulator.
- Use the same orange particle type by selecting it at the bottom of the gas pump.
- Add gas into the chamber by using the "pump" icon.
- Change the temperature of the gas chamber to about 100K by using a cooling icon at the bottom of the gas chamber.
- Click the pause button and note down the temperature, pressure, and average speed.
- Click the play button and increase the temperature of the chamber by 50K.
- Click the pause button and note down temperature, pressure, and average speed.
- Repeat the last three steps with a temperature increment of 50K till the table is completed.
- Make a graph of $\sqrt{P}$ vs average speed and discuss the behavior of graph.
- Make a graph of $\sqrt{T}$ vs average speed and discuss the behavior of graph.

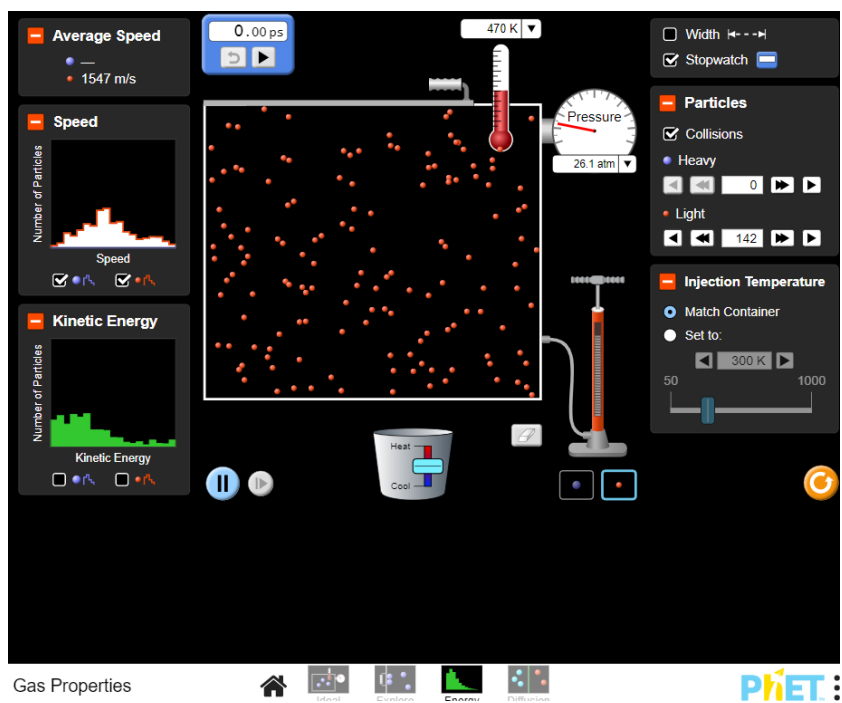

Figure 2    Simulation for analysis of average speed of gas molecules
(Picture credit: https://phet.colorado.edu/)

**PRE LAB QUESTIONS**

1) Explain Boyle's law of gas?
2) Explain Charles' law of gas?
3) Explain ideal gas law?
4) Explain what changes should be added to ideal gas law to make it applicable for real gas?
5) How the speed of gas molecules relates with the pressure and temperature of gas?

**DATA ANALYSIS AND CALCULATIONS**

*Part A: Study Boyle's law behavior of gas*

Table 1      Analysis of Boyle's law

| Width W [    ] | Volume V [    ] | Pressure P [    ] | 1/Pressure 1/P [    ] | PV [      ] |
|---|---|---|---|---|
|  |  |  |  |  |
|  |  |  |  |  |
|  |  |  |  |  |
|  |  |  |  |  |
|  |  |  |  |  |
|  |  |  |  |  |

- Make a graph of volume vs pressure and discuss the behavior of the graph.
- Make a graph of volume vs 1/pressure and discuss the behavior of the graph.

*Part B: Study Charles law behavior of gas*

Table 2      Analysis of Charles's law

| Gas type 1 Select the orange color particles | | |
|---|---|---|
| Width W [    ] | Volume V [    ] | Temperature T [    ] |
|  |  |  |
|  |  |  |
|  |  |  |
|  |  |  |
|  |  |  |
|  |  |  |

- Make a graph of volume vs temperature and discuss the behavior of graph in terms of Boyle's law.

*Part C: Study temperature vs pressure behavior of gas*

Table 3     Analysis of temperature vs pressure behavior of gas

| Gas type 1 Select the orange color particles | | |
|---|---|---|
| Temperature T [    ] | Pressure P [    ] | P/T [    ] |
| | | |
| | | |
| | | |
| | | |
| | | |
| | | |
| | | |

- Make a graph of pressure vs temperature and discuss the behavior of the graph in terms of gas laws.

*Part D: Study average speed of gas molecule as a function of temperature and pressure*

Table 4     Analysis of average speed of gas molecules

| Temperature T [    ] | $\sqrt{T}$ [    ] | Pressure P [    ] | $\sqrt{P}$ [    ] | Average speed $v_{rms}$ [    ] |
|---|---|---|---|---|
| | | | | |
| | | | | |
| | | | | |
| | | | | |
| | | | | |
| | | | | |
| | | | | |

- Make a graph of $\sqrt{P}$ vs average speed and discuss the behavior of the graph.
- Make a graph of $\sqrt{T}$ vs average speed and discuss the behavior of the graph.

# EXPERIMENT 6    REFLECTION AND REFRACTION

**OBJECTIVE**

Refraction and refraction of light rays are studied. Total internal reflection is investigated.

**THEORY AND PHYSICAL PRINCIPLES**

Visible part of the electromagnetic (EM) spectrum consists of a range of colors usually called rainbow colors as shown in figure 1. Each color has its unique wavelength ($\lambda$) and frequency ($f$) and all of them are moving as EM wave with the speed $c = 2.99 \times 10^8 \ \frac{m}{s}$ in air (free space or vacuum).

Wavelength and frequency are related as,

$$c = \lambda f \tag{1}$$

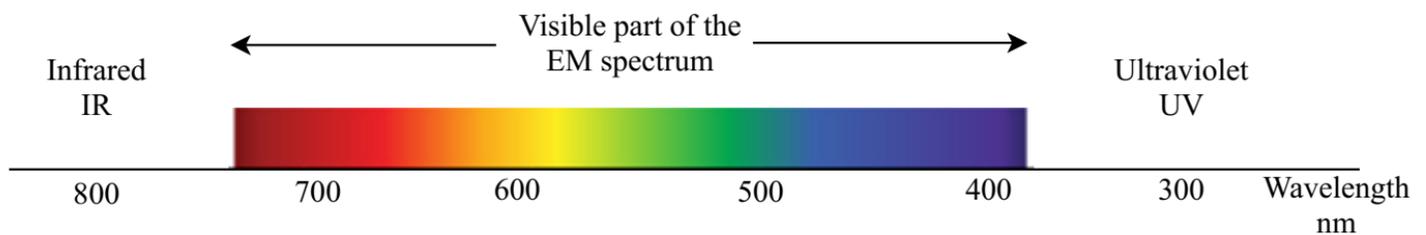

Figure 1    Visible part of the electromagnetic (EM) spectrum

When a light ray bounces back from a surface is called the reflection. A ray diagram due to reflection from a plane mirror is shown in figure 2. When a light ray is reflected from a surface, then law of reflection can be applied. Law of reflection is the incident and reflection angles should be the same. Both incident and reflected angles are measured from the normal axis at the point of incident.

$$\theta_i = \theta_r \tag{2}$$

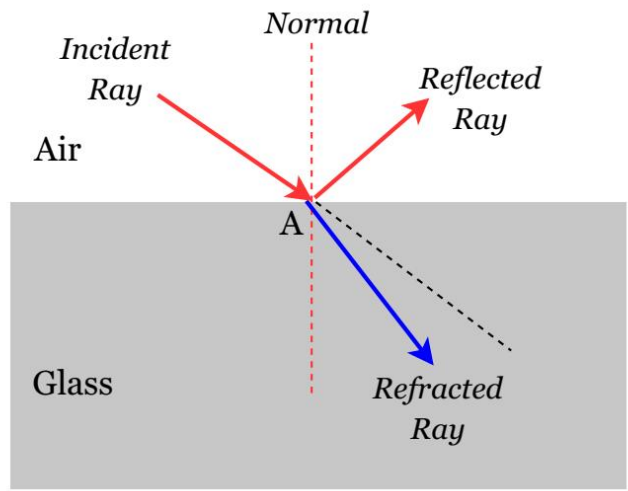

Figure 2    Reflection and refraction of light

When light ray pass through one medium to another it is called refraction. This bending nature of light (as shown in figure 2) is due to changing speed of the light rays in different mediums. Ratio of the speed of light in any medium and the speed of the light in air (or free space) is called the refractive index of light.

Refractive index,        $n = \frac{c}{v}$                                                                                     (6)

Snell law explains the relationship between incident and refracted angles and the refractive index of the medium.

$$n_1 sin\theta_i = n_2 sin\theta_r$$                                                                              (7)

Refractive index of air is one, $n_1$ =1 and the refractive index of medium is $n_2$=$n$.

$$n = \frac{sin\theta_i}{sin\theta_r}$$                                                                                        (8)

By measuring angle of incidence and angle of reflection it is possible to find the refractive index of a medium.

When light rays passing through medium of higher refractive index to medium of lower refractive index refracted ray bending away from the normal axis. This leads to a specific observation called total internal reflection which happens at one specific incident angle called critical angle.

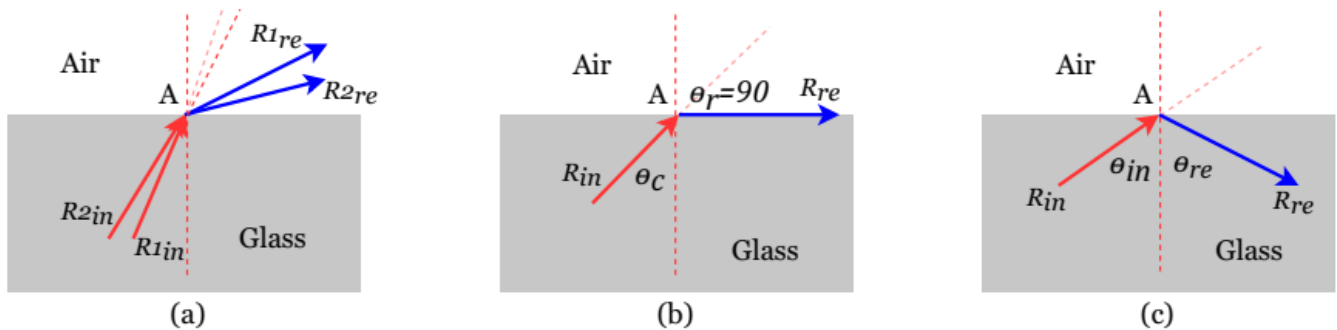

Figure 3    (a) Refraction from medium of higher refractive index to medium of lower refractive index, (b) critical incident angle, (c) total internal reflection when $\theta_{in} > \theta_C$

Apply Snell law to refraction at point A,

$$n\, sin\theta_c = sin90$$                                                                              (9)

$$n = \frac{1}{sin\theta_c}$$                                                                                        (10)

Refractive index of a medium can be found by measuring critical angle at the point of total internal reflection.

## APPARATUS AND PROCEDURE

- A very detail lesson video with detail of data collection with simulator and data analysis with excel can be found here: part 1-https://youtu.be/37os0m1Krcg and part 2- https://youtu.be/N2z08RSsX9c

*Part A: Law of reflection*

- This part of the experiment is done with following simulation: https://phet.colorado.edu/sims/html/bending-light/latest/bending-light_en.html
- Switch on the light ray.
- Set the protractor midpoint to place where the light ray hits the surface.
- Set the incident-medium to air and other-medium to water.
- Change incident angle from $10.0^0$ to $60.0^0$ and measure respective reflected angle.
- Confirm the law of reflection by comparing incident and reflection angle.
- Change the other medium into glass.
- Change incident angle from $10.0^0$ to $60.0^0$ and measure respective reflected angle.
- Confirm the law of reflection by comparing incident and reflection angle.

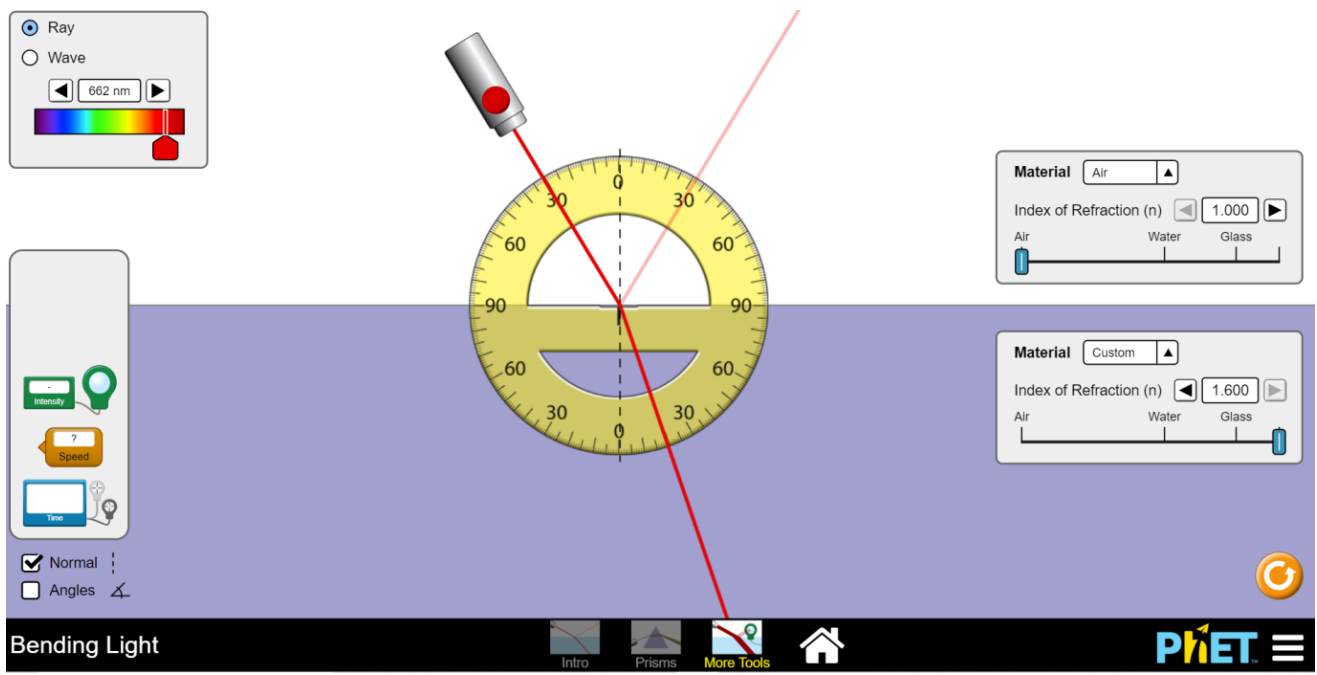

Figure 5    Simulation to invaginate laws of reflection and refraction
(Picture credit: https://phet.colorado.edu/)

*Part B: Law of refraction*

- This part of the experiment is done with the same simulation used for part B.
- In this part incident and refracted angles are measured.
- Switch on the light ray.
- Set the protractor midpoint to place where the light ray hits the surface.
- Set the incident-medium to air and refracted-medium to water.
- Change incident angle from $10.0^0$ to $60.0^0$ and measure respective refracted angles.

- Find the refractive index of water by using Snell law for each of the measured angles.
- Change the refracted-medium into glass.
- Change incident angle from $10.0^0$ to $60.0^0$ and measure respective refracted angle.
- Find the refractive index of glass by using Snell law for each of the measured angles.
- Change the refracted-medium into mystery-A.
- Change incident angle from $10.0^0$ to $60.0^0$ and measure respective refracted angle.
- Find the refractive index of the new medium by using Snell law for each of the measured angles.
- Change the refracted-medium into mystery-A.
- Change incident angle from $10.0^0$ to $60.0^0$ and measure respective refracted angle.
- Find the refractive index of the new medium by using Snell law for each of the measured angles.
- Find the refractive index of the medium by graphical method.
  - Find the sin values of incident angles.
  - Find the sin values of refracted angles.
  - Make a graph of $sin\theta_i$ vs $sin\theta_r$.
  - Fit the data with linear fitting and find the slope of the graph.
  - Find the refractive index by using the slope of the graph.

*Part C: Total internal reflection and critical angle*

- This part of the experiment is done with the same simulation used for part B.
- In this part the incident medium should have a higher refractive index to identify critical angle.
- Set the incident medium to water and refractive medium to air.
- Increase the incident angle starting from about $10^0$ and indemnify exact incident angle when the refracted angle is $90^0$.
- Find the refractive index of water by using critical angle.
- Set the incident medium to glass and refractive medium to air.
- Increase the incident angle starting from about $10^0$ and indemnify exact incident angle when the refracted angle is $90^0$.
- Find the refractive index of glass by using critical angle.
- Set the incident medium to mystery-A and refractive medium to air.
- Increase the incident angle starting from about $10^0$ and indemnify exact incident angle when the refracted angle is $90^0$.
- Find the refractive index of mystery-A by using critical angle.
- Set the incident medium to mystery-Band refractive medium to air.
- Increase the incident angle starting from about $10^0$ and indemnify exact incident angle when the refracted angle is $90^0$.
- Find the refractive index of mystery-B by using critical angle.

**PRE LAB QUESTIONS**

1) Explain reflection of light and law of reflection?
2) Explain refraction of light?
3) Explain Snell law of refraction?
4) Explain total internal reflection?
5) Name real life engineering applications of total internal reflection?

**DATA ANALYSIS AND CALCULATIONS**

*Part A: Law of reflection*

Table 1    Analysis of law of reflection

| Incident Medium = Air Other Medium = Water | | | Incident Medium = Air Other Medium = Glass | | |
|---|---|---|---|---|---|
| Angle of incident, $\theta_{in}$ [    ] | Angle of reflection, $\theta_{re}$ [    ] | PD between $\theta_{in}$ and $\theta_{re}$ [    ] | Angle of incident, $\theta_{in}$ [    ] | Angle of reflection, $\theta_{re}$ [    ] | PD between $\theta_{in}$ and $\theta_{re}$ [    ] |
| | | | | | |
| | | | | | |
| | | | | | |
| | | | | | |
| | | | | | |
| | | | | | |

*Part B: Law of refraction*

Table 2    Analysis of law of refraction

| Incident Medium = Air Refracted Medium = Water | | | Incident Medium = Air Refracted Medium = Glass | | |
|---|---|---|---|---|---|
| Incident angle $\theta_{in}$ | Refracted Angle $\theta_{re}$ | Index of Refraction $n_w(exp)$ | Incident angle $\theta_{in}$ | Refracted Angle $\theta_{re}$ | Index of Refraction $n_g(exp)$ |
| | | | | | |
| | | | | | |
| | | | | | |
| | | | | | |
| | | | | | |
| | | | | | |
| | | | | | |

Table 3    Refractive index by graphical method

| Incident Medium = Air Refracted Medium = Water | | Incident Medium = Air Refracted Medium = Glass | |
|---|---|---|---|
| Incident angle $Sin\ \theta_{in}$ | Refracted Angle $Sin\ \theta_{re}$ | Incident angle $Sin\ \theta_{in}$ | Refracted Angle $Sin\ \theta_{re}$ |
| | | | |
| | | | |
| | | | |
| | | | |
| | | | |
| | | | |
| | | | |

- Find the average refractive index of water, $n_w$(avg)?
- Compare (find percent error) between $n_w$(avg) and know value?
- Find the average refractive index of glass, $n_g$(avg)?
- Compare (find percent error) between $n_g$(avg) and know value?
- Make a graph of $sin\theta_{in}$ vs $sin\theta_{re}$.
- Fit the data with linear fitting and find the refractive index of medium $n$(graph) by using slope of the graph.
- Compare (percent difference) $n$(graph) vs $n$(avg)?

Table 4    Finding refractive index of mystery mediums by using law of refraction

| Incident Medium = Air Refracted Medium = Mystery-A | | | Incident Medium = Air Refracted Medium = Mystery-B | | |
|---|---|---|---|---|---|
| Incident angle $\theta_{in}$ | Refracted Angle $\theta_{re}$ | Index of Refraction $n_A(exp)$ | Incident angle $\theta_{in}$ | Refracted Angle $\theta_{re}$ | Index of Refraction $n_B(exp)$ |
| | | | | | |
| | | | | | |
| | | | | | |
| | | | | | |
| | | | | | |
| | | | | | |
| | | | | | |

Table 5    Refractive index by graphical method

| Incident Medium = Air Refracted Medium = Mystery-A | | Incident Medium = Air Refracted Medium = Mystery-B | |
|---|---|---|---|
| Incident angle $Sin\,\theta_{in}$ | Refracted Angle $Sin\,\theta_{re}$ | Incident angle $Sin\,\theta_{in}$ | Refracted Angle $Sin\,\theta_{re}$ |
| | | | |
| | | | |
| | | | |
| | | | |
| | | | |
| | | | |
| | | | |

- Find the average refractive index of medium mystery-A, $n_A$(avg)?
- Make a graph of $sin\theta_{in}$ vs $sin\theta_{re}$.
- Fit the data with linear fitting and find the refractive index of medium-A $n_A$(graph) by using slope of the graph.
- Compare (percent difference) $n_A$(graph) vs $n_A$(avg)?
- Find the average refractive index of medium mystery-A, $n_A$(avg)?
- Make a graph of $sin\theta_{in}$ vs $sin\theta_{re}$.
- Fit the data with linear fitting and find the refractive index of medium-A $n_A$(graph) by using slope of the graph.
- Compare (percent difference) $n_A$(graph) vs $n_A$(avg)?

*Part C: Total internal reflection and critical angle*

Table 6    Total internal reflection and critical angle

| Incident Medium | Critical angle measured $\theta_C$ [    ] | Refractive index by critical angle $n$(critical) | PD between $n$(critical) and $n$(known) |
|---|---|---|---|
| Water | | | |
| Glass | | | |
| Mystery-A | | | |
| Mystery-B | | | |

# EXPERIMENT 7    IMAGE FORMATION OF MIRRORS

## OBJECTIVE

Image formation and ray diagrams from plane, concave and convex mirrors are investigated. Focal length and center of curvature of spherical mirrors are calculated.

## THEORY AND PHYSICAL PRINCIPLES

Geometric ray diagrams are a unique method to understand the image formations of mirrors and lenses. Mirrors reflect light intensity completely back into the side of the incident light rays. When the mirror is straight it is called a plane mirror and figure 1 shows the ray diagram and image formation of the plane mirror.

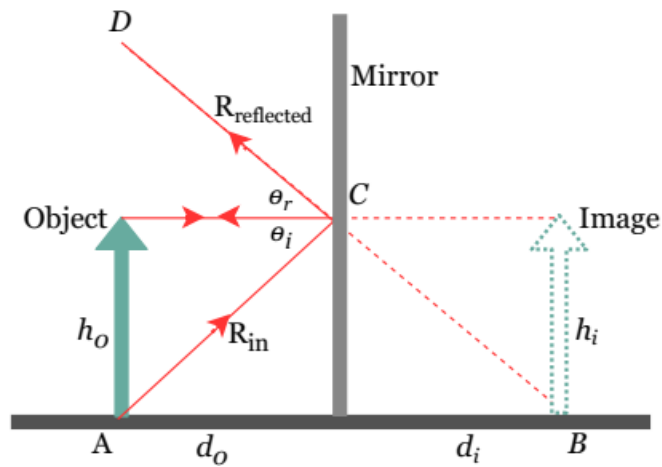

Figure 1    Plane mirror ray diagram and image formation

Geometric ray diagram of plane mirror shows that object distance ($d_0$) and image distance ($d_i$) are the same.

$$d_0 = d_i \tag{1}$$

Plane mirror ray diagram shows that image size ($h_i$) is same as the object ($h_0$). That shows the magnification of a plane mirror is one.

$$h_0 = h_i \tag{2}$$

$$m = \frac{h_i}{h_o} = \frac{-d_i}{d_o} \tag{3}$$

When measuring the object and image distances, those are measured from the mirror. In this chapter, these distances can consist of positive or negative signs. If measured position of object and/or image made by real optic ray, then it should consist of positive sign and if not or made by extrapolated rays (dotted lines) then it should consist of negative sign.

There are two types of images that can be identified: a) real image: made by only real optic rays and b) virtual image: made by only virtual (extrapolated) optic rays.

Mirrors can be made with curved surfaces. Image formation and ray diagrams of curved mirrors are quite different due to curvature of the mirrors. Only spherical mirrors and their image formation and ray diagrams (as shown in figure 2) are considered.

There are two types of mirrors, a) concave mirror: curvature towards the object and this type of mirror is converging type and b) convex mirror: curvature away from the object and this type of mirror is diverging type.

To produce a ray diagram of spherical mirrors following standard rays can be used.

1) Optic ray 1 (figure 2): parallel to optical axis → must converge to focal point of concave mirror or diverge from a focal point of convex mirror.
2) Optic ray 2 (figure 2): pass through focal point of mirror → must reflect parallel to optical axis
3) Optic ray 3 (figure 2): pass through radius of curvature → must reflect on same path as incident ray.
4) Optic ray 4 (figure 2): coming to vertex point of mirror → should reflect with same incident angle.

Figure 2(a) shows the ray diagram for image formation of a concave mirror. It can be identified that image location, size and type depend on the position of the object relative to the focal point of the concave mirror.  Figure 2(b) shows that ray diagram for image formation of convex mirror and it is identified that image from convex mirror is always smaller, upright virtual and it is not changing with the location of the object.

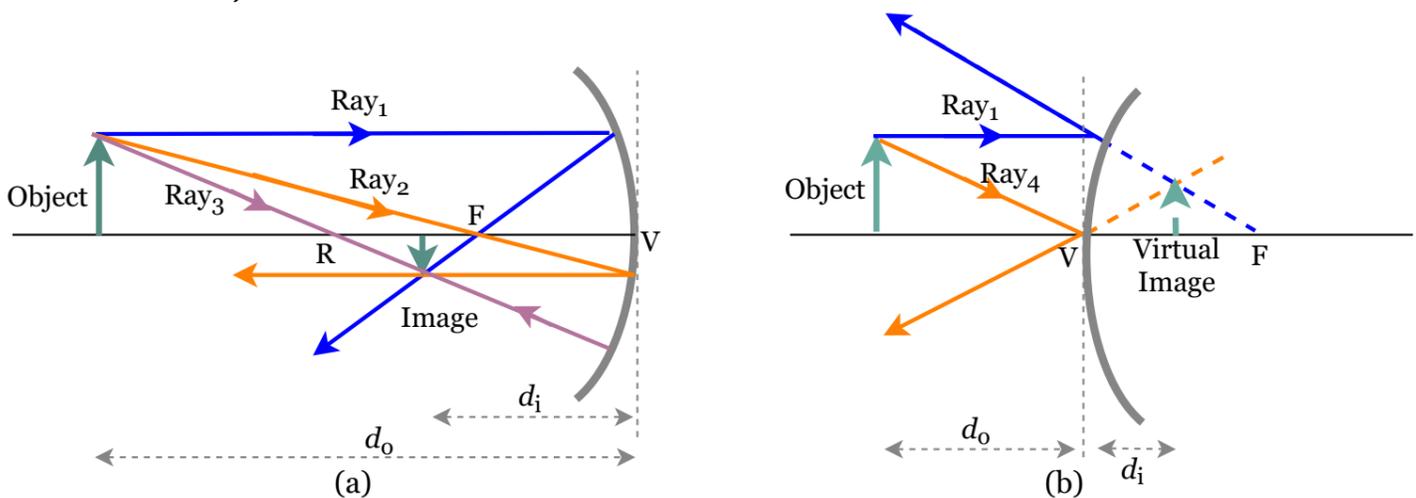

Figure 2      Ray diagram for image formation of (a) concave mirror and (b) convex mirror

A relationship between object distance ($d_0$), image distance ($d_i$) and focal point can be identified by using geometric ray diagrams for both types of mirrors and it is called mirror equation.

$$\frac{1}{d_0} + \frac{1}{d_i} = \frac{1}{f} \tag{4}$$

Table 1  Sign conventions for spherical mirrors

|  | Concave Mirror | | | Convex Mirror | | |
|---|---|---|---|---|---|---|
| $d_0$ | $f$ | $d_i$ | *Image* | $f$ | $d_i$ | *Image* |
| $> f$ | + | + | real | - | - | virtual |
| $< f$ | + | - | virtual | - | - | virtual |

## APPARATUS AND PROCEDURE

- A very detail lesson video with detail of data collection with simulator and data analysis with excel can be found here: https://youtu.be/9-QHXIEvIoY

*Part A: Ray diagram and images of plane mirror*

- This part of the experiment is done with following simulation: https://ophysics.com/l9.html
- This simulation works fine with any web browser.

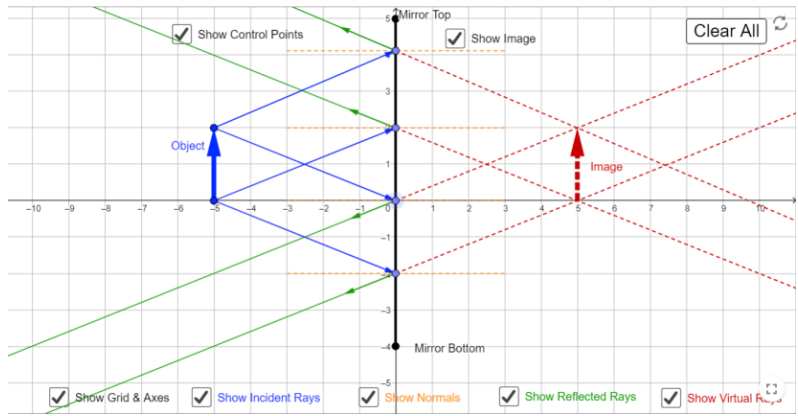

Figure 3    Simulation to invaginate images of plane mirror (Picture credit: https://ophysics.com/)

- Click on all the followings: show image, show grid and axes, show incident rays, show normal, show reflected rays, show virtual image.
- Set the object height to about two squares in grid ($h_0$=2.00cm).
- Set the object position to 2.00cm. (Negative sign in axes does not have any meaning to this experiment)
- Using one of the incident rays and its reflected ray, find the incident and reflected angles.
- Confirm the law of reflection by using incident and reflected angles.
- Change the object position by 2.00cm at a time and repeat the last two procedures.

*Part B: Draw a ray diagram and identify the virtual image of the plane mirror*

- This part of the experiment is done with following simulation: https://ricktu288.github.io/ray-optics/simulator/

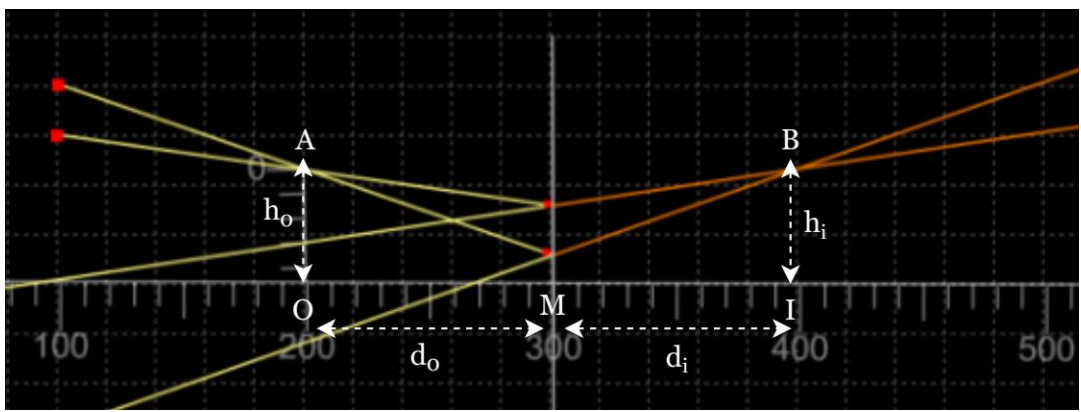

Figure 4    Ray diagram to invaginate image formation of plane mirror

- In this part of the experiment ray diagram to identify the image formation of the plane mirror is investigated.
- Click on the "grid" and use "200" in to zoom.
- Click on "ruler" and draw a horizontal ruler (as shown in the figure 4) in the middle of the grid and it can be used as the principal (optical) axis. Least count of meter ruler is 1.0 cm (10.0 mm) and numbers in the meter ruler has no meaning and those are given in pixels.
- Click on "mirror" and select "segment" and draw mirror segment (a plane mirror) on the grid as shown on the figure 4. Vertical line of "M" represents the plane mirror segment.
- Then click on "ruler" again and make an object of about a few centimeters in length. OA represents the object.
- Click on "ray" and make a ray (the red dot to the left of the object is the starting point of the ray) passing through the very top of the object and the point of intersection with the mirror is marked with a second red dot.
- Click on "extended rays" to visualize virtual rays. Yellow lines are real optical rays, and the orange lines are the virtual (extended) optical rays.
- Repeat the last two steps to make another optical ray pass through the top of the object.
- Virtual image location can be identified on the other side of the mirror when the two orange lines are intersecting.
- Click on "ruler" and measure height (IB) of the virtual image and measure the object (MO) and image (MI) distances from the mirror.

*Part C: Ray diagram and image formation of spherical mirrors and mirror equation*

- This part of the experiment is done with following simulation: https://ophysics.com/l10.html
- This simulation works fine with any web browser.

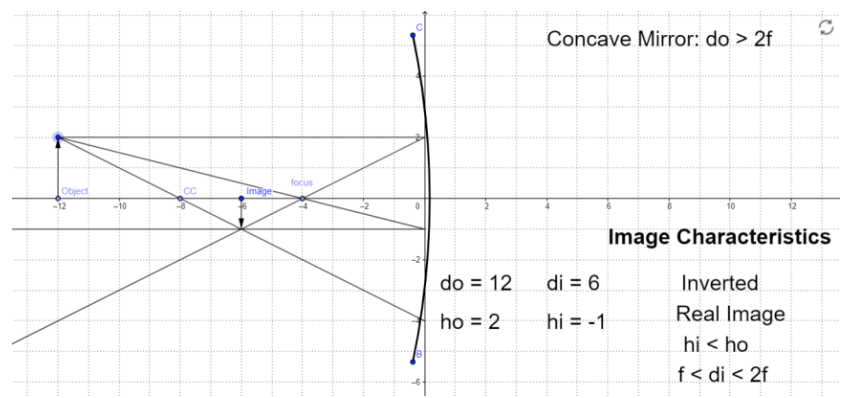

Figure 5    Simulation to invaginate images of spherical mirrors (Picture credit: https://ophysics.com/)

- Set the object height to about three squares in grid ($h_0$=3.00cm).
- Set the object position to 12.00cm on the left side of the mirror. Then the mirror acts like the concave type. (Negative sign in axes does not have any meaning to this experiment)
- Using two of the standard optic rays, the image is formed on the grid of the simulator.
- Note down the following object distance, image distance, image details (upright/inverted, real/virtual and smaller/larger).
- Repeat the last step with moving the object about 1.00cm towards the mirror (make sure $d_0$ > f).
- Calculate focal length of the mirror by using measured object and image distance and mirror equation.
- Make a graph of $1/d_0$ vs $1/d_i$ and do the trend line with linear fitting. Find the focal length of the mirror by using y intercept of the fitting.

- Set the object position to 12.00cm right side of the mirror. Then the mirror acts like convex type. (Negative sign in axes does not have any meaning to this experiment)
- Using two of the standard optic rays, the image is formed on the grid of the simulator.
- Note down the following object distance, image distance, image details (upright/inverted, real/virtual and smaller/larger).
- Repeat the last step with moving the object about 2.00cm towards the mirror.
- Calculate focal length of the mirror by using measured object and image distance and mirror equation.
- Make a graph of $1/d_0$ vs $1/d_i$ and do the trend line with linear fitting. Find the focal length of the mirror by using y intercept of the fitting.

*Part D: Draw a ray diagram and identify the images of spherical mirrors*

*Concave Mirror*

- This part of the experiment is done with following simulation: https://ricktu288.github.io/ray-optics/simulator/

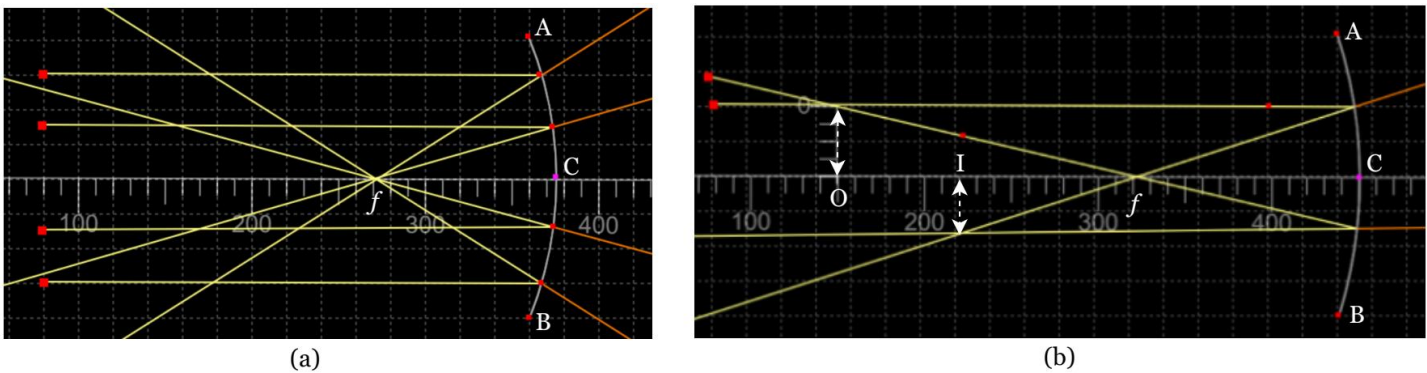

(a)                                                                        (b)

Figure 6    Ray diagram to invaginate, (a) focal point and (b) image formation of concave mirror

- In this part of the experiment ray diagram to identify the image formation of the concave mirror is investigated.
- Click on the "grid" and use "200" in to zoom.
- Click on "ruler" and draw a horizontal ruler (as shown in the figure 6) in the middle of the grid and it can be used as the principal (optical) axis. Least count of meter ruler is 1.0 cm and numbers in the meter has no meaning and those are given in pixels.
- Click on "mirror" and select "circular arc" and draw concave mirror on the grid as shown on the figure 6. To draw the concave mirror, click on point A (red dot appears) then click on point B (red dot appears) then click on point C. Point C should be to the right side and it can be adjusted to change the curvature of the mirror.

    i.    Focal point of concave mirror
- First check the focal point of the mirror by drawing parallel rays to the principal (optical) axis. Parallel rays must converge after reflection from a concave mirror.
- Click on "ray" and starting point a ray is the red dot and intersect point of ray and the mirror is the second red dot.
- Click on "extended rays" to visualize the virtual rays on the other side of the mirror.
- Take a screenshot of focal point investigation and add it to the data table section.

ii.     Image formation of concave mirror with different object location

- Continue with image formation of the concave mirror.
- Then click on "ruler" again and make an object of about two squares in the grid. O-represent the object.
- Place the object beyond the center of curvature (C = R = 2*f*) point.
- Click on "ray" and make a ray (the red dot to the left of the object is the starting point of the ray) passes through the very top of the object and parallel to the principal (optical) axis. Point of intersection with the mirror is marked with a second red dot.
- Use the second ray starts left to object and passes through top of the object and passes through the focal point of the mirror. This ray must be reflecting parallel to the principal (optical) axis.
- Image can be identified on the point where the two rays are intersecting.
- Click on "extended rays" to visualize virtual rays. Yellow lines are real optical rays, and the orange lines are the virtual (extended) optical rays.
- Take the screenshot and add it to the data table.
- Repeat all the above procedures for image formation by changing the object location to two more places (one point must be between focal point and center of curvature and the other point must be between focal point and the mirror).
- When the object is placed between the focal point and the mirror, a second optical ray must be used to pass through the top of the object and to hit at the vertex point of the mirror.
- Take screenshots of the ray diagrams and add them to the data table.

Convex Mirror

- In this part of the experiment ray diagram to identify the image formation of the convex mirror is investigated.
- Click on the "grid" and use "200" in to zoom.
- Click on "ruler" and draw a horizontal ruler (as shown in the figure 6) in the middle of the grid and it can be used as the principal (optical) axis. Least count of meter ruler is 1.0 cm and numbers in the meter ruler have no meaning and those are given in pixels.
- Click on "mirror" and select "circular arc" and draw concave mirror on the grid as shown on the figure 6. To draw the concave mirror, click on point A (red dot appears) then click on point B (red dot appears) then click on point C. Point C should be adjusted to the left side and it can be used to change the curvature of the mirror.

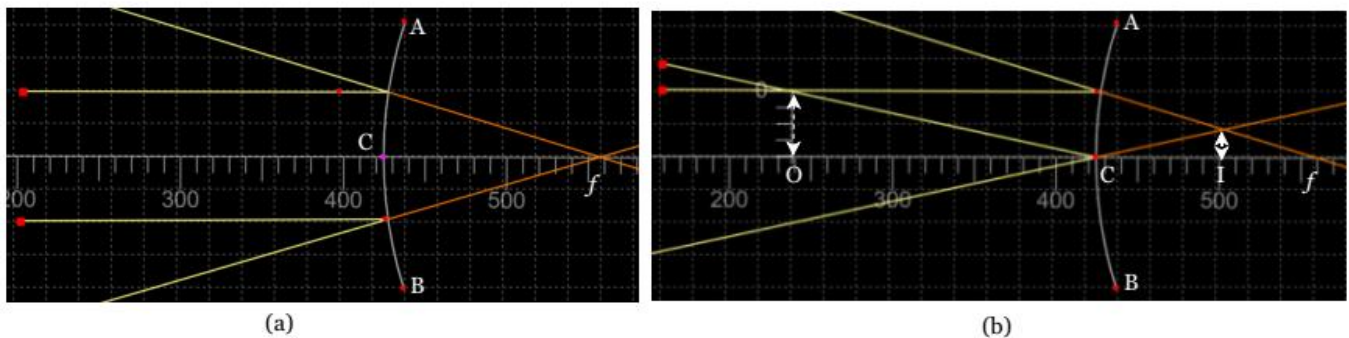

Figure 7     Ray diagram to invaginate, (a) focal point and (b) image formation of convex mirror

iii.    Focal point

- First check the focal point of the mirror my drawing parallel rays to principal axis. After reflection it should be diverged, and extended ray should pass through focal point.
- Click on "ray" and starting point a ray is the red dot and intersect point of ray and the mirror is the pink dot.
- Click on "extended rays" to visualize the virtual rays on the other side of the mirror.
- Take a screenshot of focal point investigation and add it to the data table section.

iv.    Image formation of convex mirror with different object location

- Continue with image formation of the convex mirror.
- Then click on "ruler" again and make an object of about two squares in the grid. O-represent the object.
- Place the object beyond the focal point.
- Click on "ray" and make a ray (the red dot to the left of the object is the starting point of the ray) passes through the very top of the object and parallel to the principal (optical) axis. Point of intersection with mirror is marked with a second red dot.
- Use the second ray to start left to the object and pass through the top of the object and go to the vertex point of the mirror.
- Image can be identified on the point where the two extended (virtual) rays are intersecting.
- Click on "extended rays" to visualize virtual rays. Yellow lines are real optical rays, and the orange lines are the virtual (extended) optical rays.
- Take the screenshot and add it to the data table.
- Repeat all the above procedures for image formation by changing the object location to one more place (the point must be between focal point and the mirror).
- Take screenshots of the ray diagram and add it to the data table.

**PRE LAB QUESTIONS**

1) How close does the object be from the plane mirror to get a full image of the object?
2) What is the smallest possible height of the mirror to observe the complete image of an object of height h?
3) Describe magnification of plane mirror?
4) Describe an image from a concave mirror when the object distance is larger than the focal length of the mirror?
5) Describe an image from a concave mirror when the object distance is smaller than the focal length of the mirror?

**DATA ANALYSIS AND CALCULATIONS**

*Part A: Ray diagram and images of plane mirror*

Table 1   Analysis of properties of reflection by using ray diagram of plane mirror

| Object position $d_0$ [    ] | Image distance $d_i$ [    ] | Object height $h_0$ [    ] | Image height $h_i$ [    ] | Angle of incident $\theta_{in}$ [    ] | Angle of reflection $\theta_{re}$ [    ] |
|---|---|---|---|---|---|
|  |  |  |  |  |  |
|  |  |  |  |  |  |
|  |  |  |  |  |  |
|  |  |  |  |  |  |
|  |  |  |  |  |  |
|  |  |  |  |  |  |

*Part B: Draw a ray diagram and identify the virtual image of the plane mirror*

Table 2   Analysis of properties of reflection by using ray diagram of plane mirror

| Object position $d_0$ [    ] | Image distance $d_i$ [    ] | Object height $h_0$ [    ] | Image height $h_i$ [    ] | PE of $d_0$ and $d_i$ [    ] | PE of $h_0$ and $h_i$ [    ] |
|---|---|---|---|---|---|
|  |  |  |  |  |  |

Insert the picture (screenshot) of the ray diagram.

*Part C: Ray diagram and image formation of spherical mirrors and mirror equation*

Table 3   Analysis of image formation of concave mirror

| Object position, $d_0$ [   ] | Image Distance, $d_i$ [   ] | Object height, $h_0$ [   ] | Image height, $h_i$ [   ] | *Magnification* $h_i/ h_0$ |
|---|---|---|---|---|
|  |  |  |  |  |
|  |  |  |  |  |
|  |  |  |  |  |
|  |  |  |  |  |
|  |  |  |  |  |
|  |  |  |  |  |
|  |  |  |  |  |

Table 4   Focal length analysis of concave mirror

| $1/d_0$ [   ] | $1/d_i$ [   ] | *Focal length* $f_{cal}$ [   ] |
|---|---|---|
|  |  |  |
|  |  |  |
|  |  |  |
|  |  |  |
|  |  |  |
|  |  |  |
|  |  |  |

- Find the average focal length, $f_{cal}$(avg)?
- Make a graph of $1/d_0$ vs $1/d_i$?
- Fit the data with linear fitting and find the focal length of the mirror $f_{graph}$ by using the graph.
- Compare (percent difference) $f_{graph}$ vs $f_{cal}$(avg)

Table 5    Analysis of image formation of convex mirror

| Object position, $d_0$ [    ] | Image Distance, $d_i$ [    ] | Object height, $h_0$ [    ] | Image height, $h_i$ [    ] | *Magnification* $h_i/h_0$ |
|---|---|---|---|---|
| | | | | |
| | | | | |
| | | | | |
| | | | | |
| | | | | |
| | | | | |
| | | | | |

Table 6    Focal length analysis of convex mirror

| *1/$d_0$* [    ] | *1/$d_i$* [    ] | *Focal length* $f_{cal}$ [    ] |
|---|---|---|
| | | |
| | | |
| | | |
| | | |
| | | |
| | | |
| | | |

- Find the average focal length, $f_{cal}$(avg)?
- Make a graph of $1/d_0$ vs $1/d_i$?
- Fit the data with linear fitting and find the focal length of the mirror $f_{graph}$ by using the graph.
- Compare (percent difference) $f_{graph}$ vs $f_{cal}$(avg)

*Part D: Ray diagram and image formation of the spherical mirrors*

Table 7    Ray diagrams of concave mirror

| Ray diagram of focal point | Ray diagram of object distance $d_0 > R$ |
|---|---|
|  |  |
| Ray diagram of object distance $R > d_0 > f$ | Ray diagram of object distance $f < d_0$ |
|  |  |

Table 8    Ray diagrams of convex mirror

| Ray diagram of focal point | Ray diagram of object distance $d_0 > R$ |
|---|---|
|  |  |
| Ray diagram of object distance $R > d_0 > f$ | Ray diagram of object distance $f < d_0$ |
|  |  |

# EXPERIMENT 8   IMAGE FORMATION OF LENSES

## OBJECTIVE

Image formation and ray diagrams from converging (bi-convex) and diverging (bi-concave) lenses are investigated. Focal point and image formation of lenses are investigated by drawing ray diagrams.

## THEORY AND PHYSICAL PRINCIPLES

Geometric ray diagrams are a unique method to understand the image formations of lenses. Image formation of lenses depends on refraction (bending of light) due to transparent material. Usually thin lenses (as shown in figure 1) are considered which are made of glass and consist of two sides with the same radius of curvature. When observing from the outer side of the curvature figure 1(a) is a converging type and it is a bi-convex lens and figure 1(b) diverging type and it is a bi-concave type lens.

To produce a ray diagram of a lens, following standard rays can be used.

1) Optic ray 1 (figure 1): parallel to optical axis → must converge to focal point of bi-convex lens or diverge from a focal point of bi-concave lens.
2) Optic ray 2 (figure 1): coming to vertex point of lens → should pass through without any effect.
3) Optic ray 3 (figure 1): pass through focal point of converging lens → refracted ray must be parallel to optical axis.
4) Optic ray 4 (figure 1): coming towards the focal point of diverging lens → refracted ray must be parallel to optical axis.

Converging (double convex) lens can make many different types of images due to the location of the object relative to the focal point. Diverging (double concave) type lenses always make a smaller, upright and virtual image and it is not depending on the location of the object.

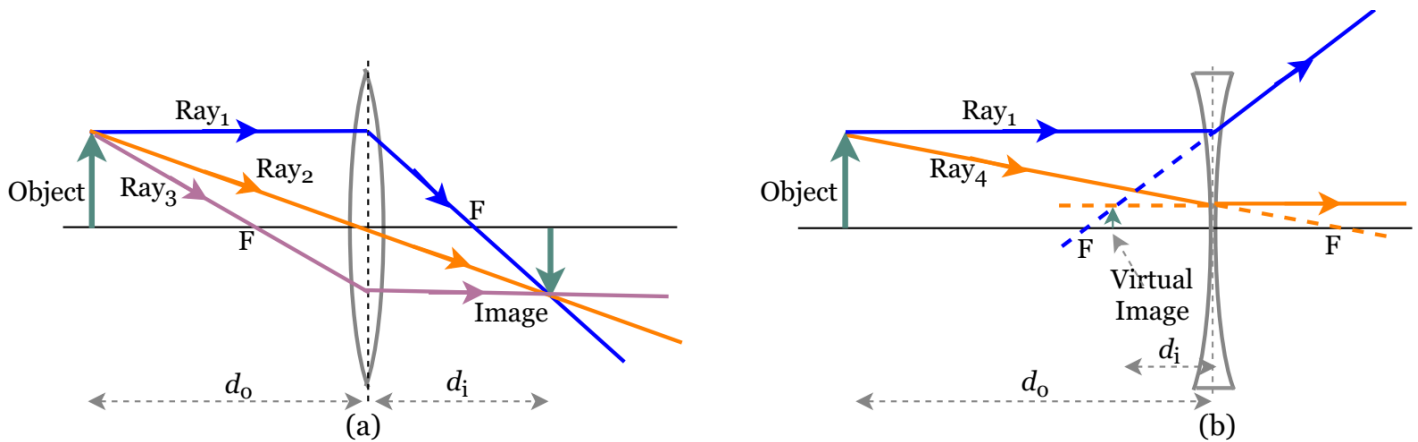

Figure 1    Ray diagrams for image formation of (a) converging (bi-convex) lens and
(b) diverging (bi-concave) lens

A relationship between object distance ($d_0$), image distance ($d_i$) and focal point can be identified by using geometric ray diagrams for both types of lenses and it is called lens equation.

$$\frac{1}{d_0} + \frac{1}{d_i} = \frac{1}{f} \tag{1}$$

Magnification equation, $m = \frac{h_i}{h_o} = \frac{-d_i}{d_o}$                                                (2)

Signs of focal point, image distance, image type relative to object position are summarized in table below.

Table 1  Sign convention of lenses

| $d_0$ | Converging (bi-convex) Lens | | | Diverging (bi-concave) Lens | | |
|---|---|---|---|---|---|---|
| | $f$ | $d_i$ | *Image* | $f$ | $d_i$ | *Image* |
| $> f$ | + | + | real | - | - | virtual |
| $< f$ | + | - | virtual | - | - | virtual |

## APPARATUS AND PROCEDURE

- A very detail lesson video with detail of data collection with simulator and data analysis with excel can be found here: https://youtu.be/_9QZzR8D1ko

*Part A: Focal point and image formation of double convex (converging) lens*

- This experiment is done with following simulation:
  https://ricktu288.github.io/ray-optics/simulator/

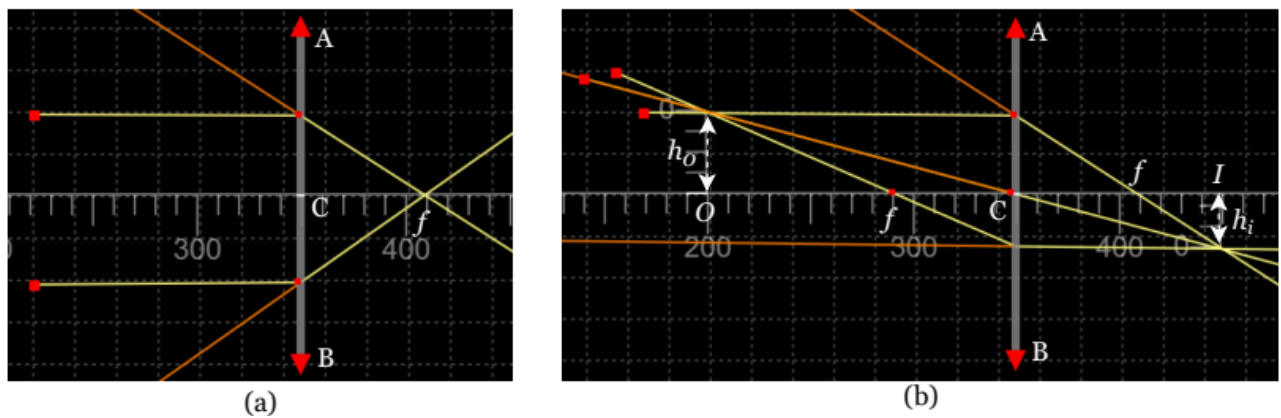

Figure 2    Ray diagram to investigate, (a) focal point and (b) image formation of double convex (converging) lens

- In this part of the experiment ray diagram to identify the image formation of the concave mirror is investigated.
- Click on the "grid" and use "200" in to zoom.
- Click on "ruler" and draw a horizontal ruler (as shown in the figure 2) in the middle of the grid and it can be used as the principal axis. Least count of meter ruler is 1.0 cm and numbers in the meter ruler have no meaning and those are given in pixels.
- Click on "glasses" and select "ideal lens" and draw the lens on the grid as shown on figure 2. To draw the lens, click on point A (red arrow) then click on another point (second red arrow).
- A lens appears and to make sure the lens is convex (converging) select and adjust focal length of the lens to any positive value.
- Click on the lens and an icon will appear on the top left of the lens to change the focal length of the lens. Type 60 in the focal length then the focal length is 6.0 cm in meter ruler measurement.

   v.     Focal point of convex (converging) lens
- First to check the focal point of the mirror optical rays must draw parallel to the principal axis. Parallel rays must converge after refracted from converging lens.
- Click on "ray" and starting point a ray is the red dot and intersect point of ray and the lens is the other red dot.
- Use at least two parallel rays (parallel to the principal axis) coming from the left side to find out the focal point on the right side of the lens.
- Click on "extended rays" to visualize the virtual rays.
- Repeat the last two steps by drawing two rays from the right side to find out the focal point on the left side.
- Take a screenshot of focal point investigation and add it to the data table section.

   vi.    Image formation of convex (converging) lens with different object location
- Continue with image formation of the convex (converging) lens.
- Then click on "ruler" again and make an object of about two squares in grid (about 4 units in meter ruler, 4.0 cm). O-represent the object.
- Place the object beyond the center of curvature (C = R = 2$f$) point.
- Click on "ray" and make a ray (the red dot to the left of the object is the starting point of the ray) passes through the very top of the object and parallel to the principal axis. Point of intersection with the lens is marked with a second red dot.
- Use the second ray which starts left to the object and passes through the top of the object and pass through the focal point (or vertex point) of the lens. This ray must be refracting parallel to the principal axis.
- Image can be identified on the point where the two rays are intersecting.
- Click on "extended rays" to visualize any virtual rays. Yellow lines are real optical rays, and the orange lines are the virtual (extended) optical rays.
- Take the screenshot and add it to the data table. Measure object and image distances and heights. Record them in the table.
- Repeat all the above procedures for image formation by changing the object location to five more places (two points must be between focal point and center of curvature and the two points must be between focal point and the lens).
- When the object is placed between the focal point and the lens, a second optical ray must be used to pass through the top of the object and to hit at the vertex point of the lens.
- Take screenshots of the ray diagrams and add them to the data table. Measure object and image distances and heights. Record them in the table.
- Calculate focal length of the lens by using measured object and image distance and lens equation.
- Calculate the magnification by using object and image distances and their heights and compare them.
- Make a graph of $1/d_0$ vs $1/d_i$ and do the trend line with linear fitting. Find the focal length of the lens by using y intercept of the fitting.
- Compare the focal length from the graph with known value.

*Part B: Focal point and image formation of double concave (diverging) lens*
- In this part of the experiment ray diagram to identify the image formation of the double concave (diverging) lens is investigated.
- Click on the "grid" and use "200" in to zoom.
- Click on "ruler" and draw a horizontal ruler (as shown in the figure 3) in the middle of the grid and it can be used as the principal axis.

- Least count of meter ruler is 1.0 cm and numbers in the meter ruler have no meaning and those are given in pixels.
- Click on "glasses" and select "ideal lens" and draw the lens on the grid as shown on figure 3. To draw the lens, click on point A (red arrow) then click on another point (second red arrow).
- A lens appears and to make sure the lens is concave (diverging) select and adjust focal length of the lens to any negative value.

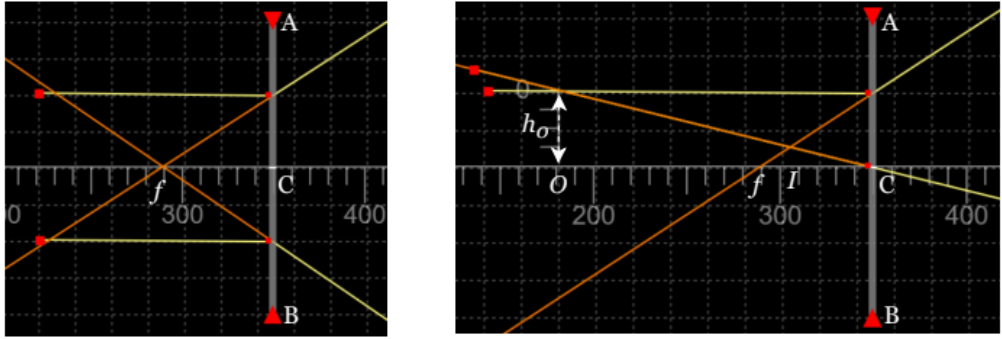

Figure 3   Ray diagram to investigate, (a) focal point and (b) image formation of double concave (diverging) lens

i.      Focal point of concave (diverging) lens
- First to check the focal point of the lens optical rays must draw parallel to the principal axis. After refraction it should be diverged, and extended ray should pass through the focal point.
- Click on "ray" and starting point a ray is the red dot and intersect point of ray and the lens is the other red dot.
- Use at least two parallel rays (parallel to the principal axis) coming from the left side to find out the focal point on the right side of the lens.
- Click on "extended rays" to visualize the virtual rays.
- Repeat the last two steps by drawing two rays from the right side to find out the focal point on the left side.
- Click on "extended rays" to visualize the virtual rays.
- Take a screenshot of focal point investigation and add it to the data table section.

ii.     Image formation of concave (diverging) lens with different object location
- Continue with image formation of the concave (diverging) lens.
- Then click on "ruler" again and make an object of about two squares in the grid. O-represent the object.
- Place the object beyond the focal point.
- Click on "ray" and make a ray (the red dot to the left of the object is the starting point of the ray) passes through the very top of the object and parallel to the principal axis. Point of intersection with the lens is marked with a second red dot.
- Use the second ray which starts left to the object and passes through the top of the object and goes to the vertex point of the lens.
- Image can be identified on the point where the two extended (virtual) rays are intersecting.
- Click on "extended rays" to visualize virtual rays. Yellow lines are real optical rays, and the orange lines are the virtual (extended) optical rays.
- Take the screenshot and add it to the data table. Measure object and image distances and heights. Record them in the table.

- Repeat all the above procedures for image formation by changing the object location to five more places (two points must be between focal point and center of curvature and the two points must be between focal point and the lens).
- Calculate focal length of the lens by using measured object and image distance and lens equation.
- Calculate the magnification by using object and image distances and their heights and compare them.
- Make a graph of $1/d_0$ vs $1/d_i$ and do the trend line with linear fitting. Find the focal length of the lens by using y intercept of the fitting.
- Compare the focal length from the graph with known value.

*Part C: Ray diagram of lens combination with object is very closer to first lens (similar to compound microscope)*

- This part of the experiment is done with following simulation:
  https://ricktu288.github.io/ray-optics/simulator/

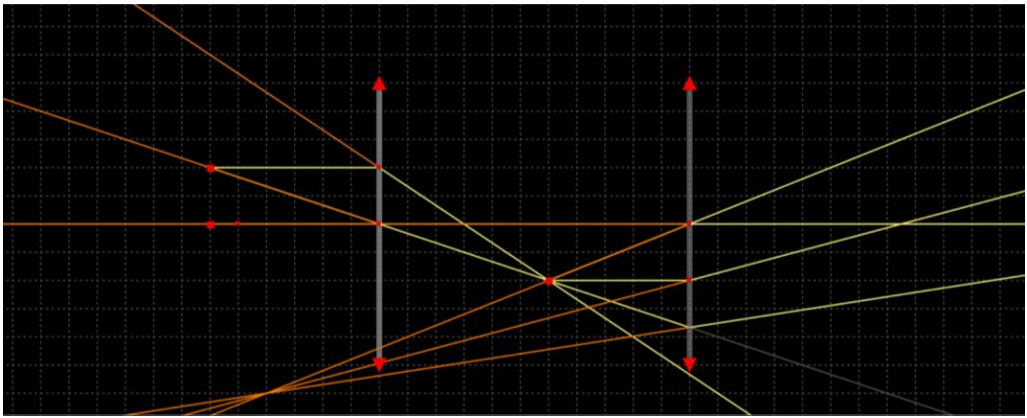

Figure 4    Simulation to investigate lens combination

- Click on "grid" and change the zoom to 200 level.
- Click on "Glasses" and select the "ideal lens".
- Draw lens 1 (objective lens) as in the figure 4 and change "focal length" to 60 in the selection panel. The focal length of an objective lens is 6.00cm ($f_0$ = 6.00cm).
- Draw lens 2 (eye piece) about 9 squares from lens 1 and change "focal length" to 100 in the selection panel. The focal length of an objective lens is 10.0cm ($f_0$ = 10.0cm).
- Click on "ray" in the selection tool and draw one ray starting from the 8th square line to the right of the objective lens. Then object distance $d_0$=14.00cm. And adjust the first ray perpendicular to the axis of the lens. It should be passed through both lenses without any bending and it can be considered as the principal axis of ray diagram.
- Create a second ray starting from the same starting point as the second ray and adjust it to pass through the vertex point of the lens. So, it should pass through the objective lens without any bending.
- Now you can identify the first image from the objective lens in between the two lenses.
- First image is the object for the second lens (eye piece) therefore objective distance for lens 2 is $d_0^/ = D - d_i$, D is the separation between lenses and $d_i$ is the image distance of lens 1.
- Then draw a new ray starting from the first image location and adjust it to pass through the eyepiece with parallel to the principal axis.
- Create another ray starting from the first image location and adjust it to pass through the vertex point of the eye piece.
- Then click on the "extended ray" icon to identify the virtual image from the eyepiece.

- Extended rays from the first image will intersect the other side of the lens 1 which is the final image of the lens combination (compound microscope).
- Add the picture of the ray diagram to the data table.

*Part D: Ray diagram of lens combination with object very far away (at infinity) to first lens (similar to simple telescope)*

- This part of the experiment is done with following simulation:
  https://ricktu288.github.io/ray-optics/simulator/

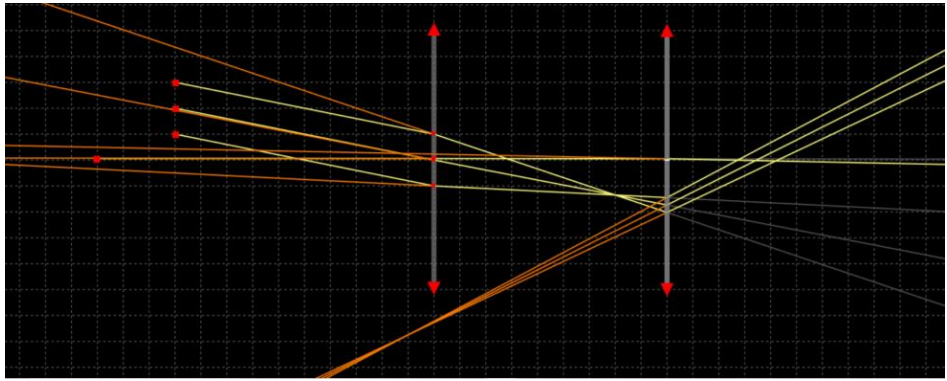

Figure 5  Simulation to investigate lens combination

- Click on "grid" and change the zoom to 200 level.
- Click on "Glasses" and select the "ideal lens".
- Draw lens 1 (objective lens) as in the figure 4 and change "focal length" to 140 in the selection panel. The focal length of an objective lens is 14.0cm ($f_0$ = 14.0cm).
- Draw lens 2 (eye piece) about 9 squares from lens 1 and change "focal length" to 50 in the selection panel. The focal length of an objective lens is 5.0cm ($f_0$ = 5.0cm).
- Click on "ray" in the selection tool and draw three parallel rays which aim at the objective lens as shown in figure 5.
- Now you can identify the first image from the objective lens in between the two lenses.
- First image is the object for the second lens (eye piece) therefore objective distance for lens 2 is $d_0^/ = D - d_i$, D is the separation between lenses and $d_i$ is the image distance of lens 1.
- Then draw a new ray starting from the first image location and adjust it to pass through the eyepiece with parallel to the principal axis.
- Create another ray starting from the first image location and adjust it to pass through the vertex point of the eye piece.
- Then click on the "extended ray" icon to identify the virtual image from the eyepiece.
- Extended rays from the first image will intersect the other side of the lens 1 which is the final image of the lens combination (compound microscope).
- Add the picture of the ray diagram to the data table.

**PRE LAB QUESTIONS**

1) Describe bi-convex (converging) lens?
2) Describe bi-concave (diverging) lens?
3) Describe the focal point properties of converging lens?
4) Describe the focal point properties of diverging lens?

## DATA ANALYSIS AND CALCULATIONS

*Part A: Ray diagram and image formation of lens and lens equation*

Table 1    Focal length and magnification of convex (converging) lens

| Object position, $d_0$ [    ] | Image distance $d_i$ [    ] | *Focal length $f_{cal}$* [    ] | Image height, $d_i$ [    ] | Magnification $m_h=h_i/h_o$ | Magnification $m_d=-d_i/d_o$ | PE of magnifications |
|---|---|---|---|---|---|---|
|   |   |   |   |   |   |   |
|   |   |   |   |   |   |   |
|   |   |   |   |   |   |   |
|   |   |   |   |   |   |   |
|   |   |   |   |   |   |   |
|   |   |   |   |   |   |   |

- Find the average focal length, $f_{cal}$(avg)?
- Make a graph of $1/d_0$ vs $1/d_i$?
- Fit the data with linear fitting and find the focal length of the lens $f_{graph}$ by using the graph.
- Compare (percent difference) $f_{graph}$ vs $f_{cal}$(avg)

Table 2   Ray diagrams of convex (converging) lens

| Focal point | |
|---|---|
| | |
| $d_0 > 2f$ | $d_0 > 2f$ |
| | |
| $2f> d_0 > f$ | $2f> d_0 > f$ |
| | |
| $f > d_0$ | $f > d_0$ |
| | |

Table 3    Focal length and image magnification of concave (diverging) lens

| Object position, $d_0$ [    ] | Image distance $d_i$ [    ] | *Focal length* $f_{cal}$ [    ] | Image height, $d_i$ [    ] | Magnification $m_h = h_i/h_o$ | Magnification $m_d = -d_i/d_o$ | PE of magnifications |
|---|---|---|---|---|---|---|
|  |  |  |  |  |  |  |
|  |  |  |  |  |  |  |
|  |  |  |  |  |  |  |
|  |  |  |  |  |  |  |
|  |  |  |  |  |  |  |
|  |  |  |  |  |  |  |

- Find the average focal length, $f_{cal}$(avg)?
- Make a graph of $1/d_0$ vs $1/d_i$?
- Fit the data with linear fitting and find the focal length of the lens $f_{graph}$ by using the graph.
- Compare (percent difference) $f_{graph}$ vs $f_{cal}$(avg)

Table 4    Ray diagrams of concave (diverging) lens

| Focal point | |
|---|---|
|  |  |
| $d_0 > 2f$ | $d_0 > 2f$ |
|  |  |
| $2f > d_0 > f$ | $2f > d_0 > f$ |
|  |  |
| $f > d_0$ | $f > d_0$ |
|  |  |

*Part C: Ray diagram and image formation of the combination of spherical lenses*

Table 5    Ray diagrams of lens combinations

| Ray diagram when the object is very close to lens 1 (objective lens), which simulates the compound microscope. |
|---|
| 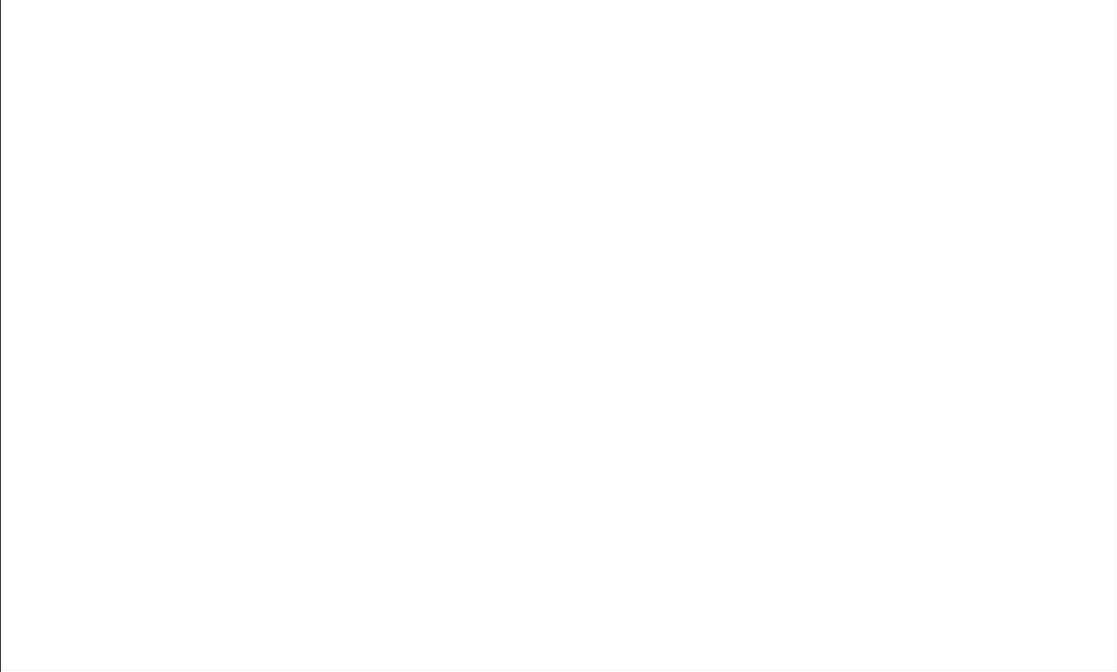 |
| Ray diagram when the object is far away (at infinity) to lens 1 (objective lens), which simulate the simple telescope. |
| 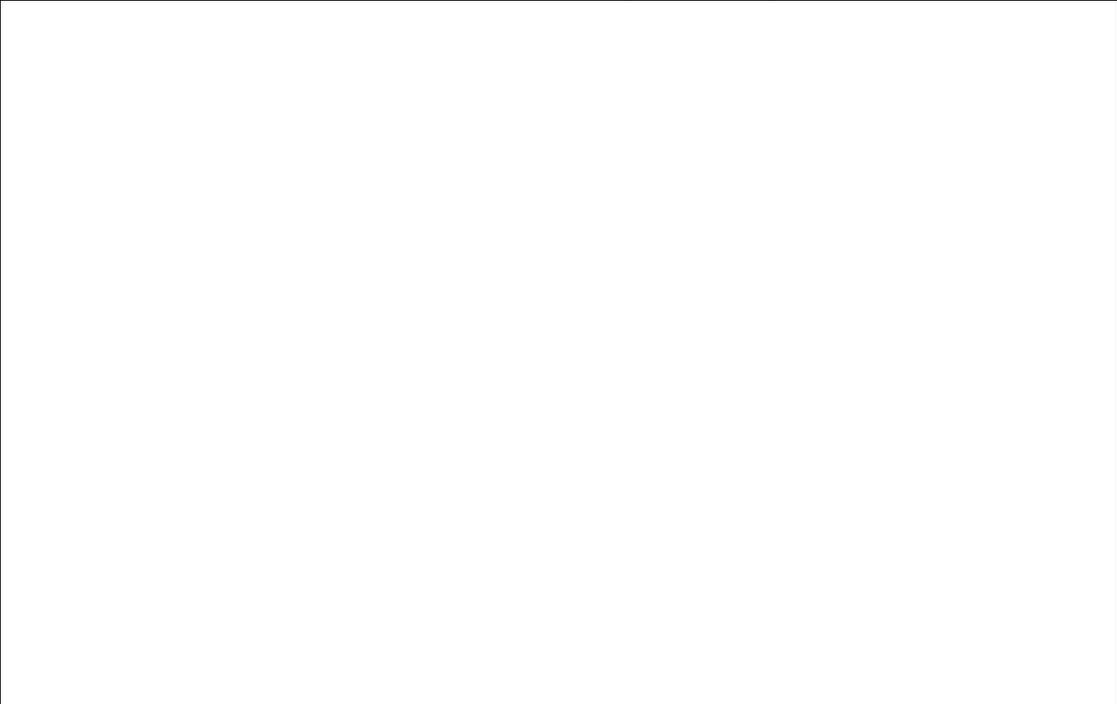 |

# EXPERIMENT 9
# INTERFERENCE AND DOUBLE SLIT EXPERIMENT

## OBJECTIVE

Interference due to two light sources is investigated. Wavelength of the light is analyzed by varying the separation between slits and screen and slit separation. Wavelengths of range of colors are investigated by using a double slit experiment.

## THEORY AND PHYSICAL PRINCIPLES

In the seventeenth century, Dutch physicist Christiaan Huygens (1629-1695) explained the bending nature of light by using Huygens principle. In which he explained that light was a wave, and he used the idea of wavelet and wavefront propagation to explain the bending nature. Huygens explanation of wave nature of light was not generally accepted till it was confirmed experimentally by double slit experiment which was done by Thomas Young (1773-1829). Figure 1 shows the schematic diagram of Young's double slit experiment and it can be observed that the two slits $S_1$ and $S_2$ act like two sources which emit wave fronts. The waves from two slits interfere and produce an interference pattern on the screen.

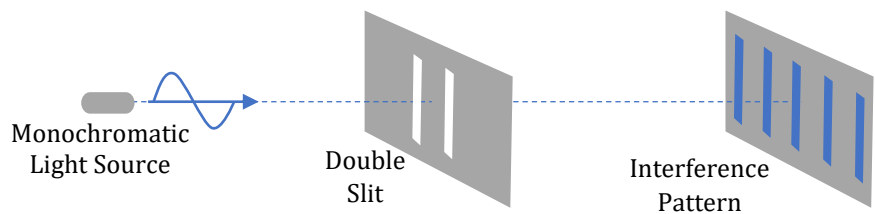

Figure 1     Schematic diagram of double slit experiment

Waves start at slit-1 and slit-2 produce secondary wave fronts therefore waves double slits are in phase or have definite phase relationship. These waves are called coherent waves. If the waves do not have a definite phase relationship, then it is called incoherent and an example of such waves are the waves from two independent light sources.

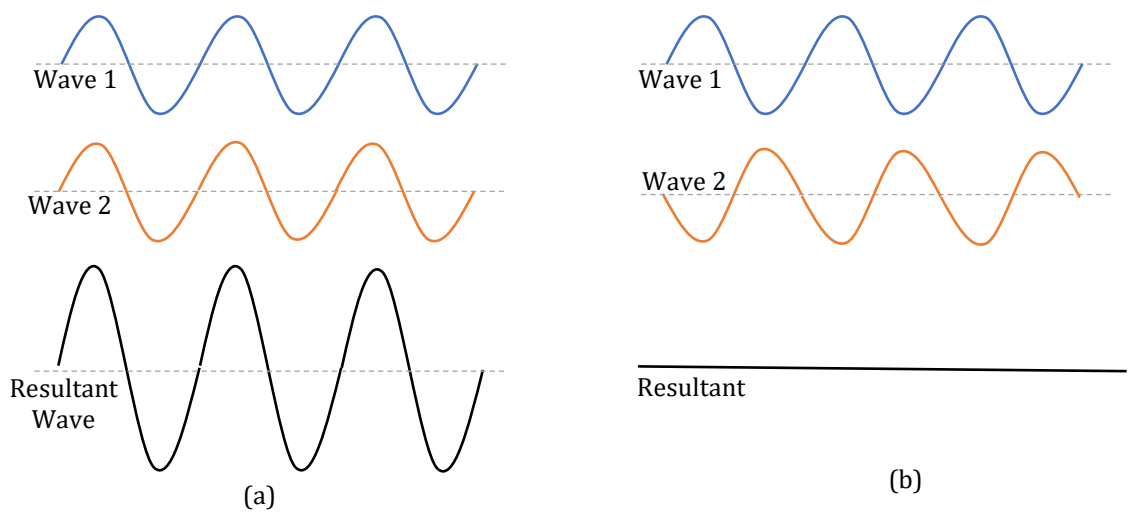

Figure 2     Schematic diagram of (a) constructive and (b) destructive interferences

Interference of waves from slit-1 and slit-2 produce two extreme intensity points on the screen as shown in the figure 1. These can be explained as constructive interference or bright fringe (see figure 2(a)) and destructive interference or dark fringe (see figure 2(b)). Bright fringes are due to interference of exact in phase waves and dark fringes are due to interference of exact out of phase waves.

To understand the interference pattern shown in the screen in figure 1, it is important to check the path length difference in two waves when they arrive at the screen which is shown in figure 3. The resultant intensity at point P located on the screen depends on the path length difference between two waves.

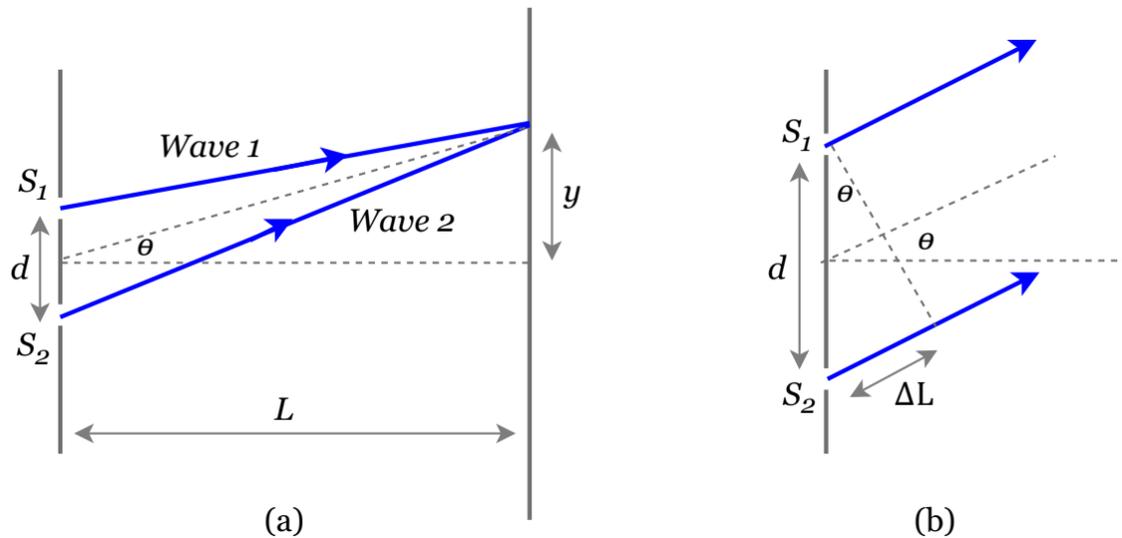

(a)                                                          (b)

Figure 3      (a) path length of two waves which starts at two slits and interfere on the screen, (b) screen is further away from slits ($d <<< L$) therefore waves looks parallel at very closer to slits

Path length difference of two waves,

$$\Delta L = dsin\theta = n\lambda \tag{1}$$

If $n$ is any integer, then it produces a bright fringe interference. For constructive or bright interference,

$$\Delta L = dsin\theta = n\lambda, \ n = 0, \pm1, \pm2, \pm3, \ldots\ldots. \tag{2}$$

$$sin\theta = \frac{y}{D} \tag{3}$$

$$d\frac{y}{D} = n\lambda \tag{4}$$

Location of a $n^{\text{th}}$ bright fringe relative to central bright fringe,

$$y_n = \frac{n\lambda D}{d} \tag{5}$$

If $n$ is any multiplier of half, then it produces a dark fringe interference. For destructive or dark interference,

$$\Delta L = dsin\theta = n\lambda, \ n = \pm\frac{1}{2}, \pm\frac{3}{2}, \pm\frac{5}{2}, \ldots\ldots. \tag{3}$$

## APPARATUS AND PROCEDURE

- This experiment is done with following simulation:
  https://phet.colorado.edu/sims/html/wave-interference/latest/wave-interference_en.html

- A very detail lesson video with detail of data collection with simulator and data analysis with excel can be found here: https://youtu.be/-M_Brj5Bugg

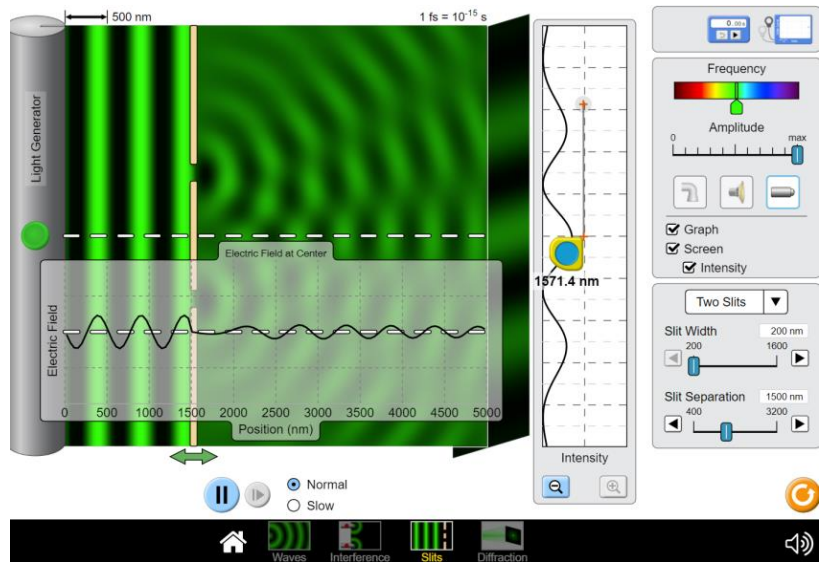

Figure 4    Simulation to investigate interference and diffraction
(Picture credit: https://phet.colorado.edu/)

*Part A: Wavelength of light color by screen distance of double slit experiment*

- Select the double slit experiment tool in the selection tools at the very bottom of the page.
- Set the frequency selector into mid of the green color.
- Set the amplitude selector into maximum.
- Switch on graph, screen, and intensity in the selection tools on the right side.
- Select the two slits on the bottom right side of the selection panel.
- Set the slit width into 200.0nm.
- Set the slit separation to 1000.0nm.
- Set the position of the slits 2000.0nm which means the separation between screen and the slits is 3000.0nm.
- Switch on the light generator on the left side of the simulator.
- Observe the interference pattern on screen.
- Measure the separation using a measurement tool between the middle of the central bright fringe and the center of the first bright fringe on screen.
- Bring the slits 500.0nm towards the screen at a time and measure the new position of the first bright fringe on screen.
- Make a graph between the position of the first bright fringe on screen and the separation between slits and the screen.
- Fit the data with linear fitting and find the wavelength of the light by using slope of the graph.
- Compare the wavelength of the light from the graph with the actual wavelength.
- Actual wavelength of the light can be measured by using the tape on the graph.
- Repeat above procedure for red and blue colors.

*Part B: Wavelength of light by using double slit fringe pattern*

- Set the slit width into 200.0nm.
- Set the slit separation to 2000.0nm.
- Set the position of the slits 2000.0nm which means the separation between screen and the slits is 3000.0nm.
- Measure the separation using a measurement tool between the middle of the central bright fringe and the center of the first bright fringe on screen.
- Measure wavelength of the light by using a graph.
- Calculate the wavelength of the light by using a double slit equation.
- Compare the wavelength values.
- Change the color of the light in table 2 by using frequency selector and repeat the above procedure.

*Part C: Wavelength of green color by varying slit separation of double slit experiment*

- Set the frequency selector into mid of the green color.
- Set the position of the slits 2000.0nm which means the separation between screen and the slits is 3000.0nm. Do not change this during this part of the experiment.
- Set the slit width into 200.0nm.
- Set the slit separation to 1000.0nm.
- Measure the separation using a measurement tool between the middle of the central bright fringe and the center of the first bright fringe on screen.
- Increase the slit separation by 500.0nm at a time and measure the new position of the first bright fringe on screen.
- Make a graph between the position of the first bright fringe on screen and the slit separation and discuss the behavior of the graph.
- Make a graph between the position of the first bright fringe on screen vs the inverse of slit separation (1/d).
- Fit the data with linear fitting and find the wavelength of the light by using slope of the graph.
- Compare the wavelength of the light from the graph with the actual wavelength.

**PRE LAB QUESTIONS**

1) Describe the bending nature of light?
2) Describe Huygens's principle?
3) Describe Young's double slit experiment?
4) Describe single slit diffraction?
5) Describe how to find the separation between two stars (stars are many light years further from Earth) by using interference patterns observed on Earth?
6) Describe interference patterns from a soap bubble?

## DATA ANALYSIS AND CALCULATIONS

*Part A: Wavelength of light by varying the location of the slits of double slit experiment*

Table 1    Analysis of light wavelength of green color

| Slit separation d = | | Wavelength measured, $\lambda_{measured}$ = | |
|---|---|---|---|
| Separation between slits and screen $D$ [     ] | Position of first bright fringe $y$ [     ] | Wavelength calculated $\lambda_{cal}$ [     ] | PD between $\lambda_{cal}$ and $\lambda_{measured}$ [     ] |
|  |  |  |  |
|  |  |  |  |
|  |  |  |  |
|  |  |  |  |
|  |  |  |  |

- Find the calculated average wavelength λcal(avg) of the light?
- Compare (percent difference) λcal(avg) with actual wavelength?
- Make a graph between the position of the first bright fringe on screen and the separation between slits and the screen. Fit the data with linear fitting and find the frequency of the light by using slope of the graph.
- Compare the wavelength of the light from the graph with the actual wavelength.

Table 2    Analysis of light wavelength of blue color

| Slit separation d = | | Wavelength measured $\lambda_{measured}$ = | |
|---|---|---|---|
| Separation between slits and screen $D$ [     ] | Position of first bright fringe $y$ [     ] | Wavelength calculated $\lambda_{cal}$ [     ] | PD between $\lambda_{cal}$ and $\lambda_{measured}$ [     ] |
|  |  |  |  |
|  |  |  |  |
|  |  |  |  |
|  |  |  |  |
|  |  |  |  |
|  |  |  |  |

- Find the calculated average wavelength λ$_{cal}$(avg) of the light?
- Compare (percent difference) λ$_{cal}$(avg) with actual wavelength?
- Make a graph between the position of the first bright fringe on screen and the separation between slits and the screen. Fit the data with linear fitting and find the frequency of the light by using slope of the graph.
- Compare the wavelength of the light from the graph with the actual wavelength.

Table 3    Analysis of light wavelength of red color

| Slit separation, d = | | Wavelength measured, $\lambda_{measured}$ = | |
|---|---|---|---|
| Separation between slits and screen $D$ [     ] | Position of first bright fringe $y$ [     ] | Wavelength calculated $\lambda_{cal}$ [     ] | PD between $\lambda_{cal}$ and $\lambda_{measured}$ [     ] |
|  |  |  |  |
|  |  |  |  |
|  |  |  |  |
|  |  |  |  |
|  |  |  |  |
|  |  |  |  |

- Find the calculated average wavelength λ$_{cal}$(avg) of the light?
- Compare (percent difference) λ$_{cal}$(avg) with actual wavelength?
- Make a graph between the position of the first bright fringe on screen and the separation between slits and the screen. Fit the data with linear fitting and find the frequency of the light by using slope of the graph.
- Compare the wavelength of the light from the graph with the actual wavelength.

*Part B: Wavelength of light by using double slit fringe pattern*

Table 4    Analysis of light wavelength of different colors with double slit experiment

| Color | Position of first bright fringe, $y$ [     ] | Wavelength calculated, $\lambda_{cal}$ [     ] | Wavelength measured, $\lambda_{measured}$ [     ] | PD between $\lambda_{cal}$ and $\lambda_{meas}$ [     ] |
|---|---|---|---|---|
| Dark Red |  |  |  |  |
| Orange |  |  |  |  |
| Yellow |  |  |  |  |
| Light Green |  |  |  |  |
| Dark Blue |  |  |  |  |
| Violet |  |  |  |  |

*Part C: Wavelength of light by varying slit separation of double slit experiment*

Table 5   Analysis of light wavelength by varying slit separation of double slit experiment

| Screen distance D = | | Wavelength measured $\lambda_{measured}$ = | | |
|---|---|---|---|---|
| Slit separation of the slits, $d$ [    ] | Position of first bright fringe, $y$ [    ] | Wavelength calculated, $\lambda_{cal}$ [    ] | $1/d$ [    ] | PD between $\lambda_{cal}$ and $\lambda_{meas}$ [    ] |
|  |  |  |  |  |
|  |  |  |  |  |
|  |  |  |  |  |
|  |  |  |  |  |
|  |  |  |  |  |

- Find the calculated average wavelength λ$_{cal}$(avg) of the light?
- Compare (percent difference) λ$_{cal}$(avg) with actual wavelength?
- Make a graph between the position of the first bright fringe on screen and the slit separation and discuss the behavior of the graph.
- Make a graph between the position of the first bright fringe on screen vs the inverse of slit separation (1/d).
- Fit the data with linear fitting and find the wavelength of the light by using slope of the graph.
- Compare the wavelength of the light from the graph with the actual wavelength.

# EXPERIMENT 10    PHOTOELECTRIC EFFECT

## OBJECTIVE

Physical characteristics of photoelectric effect are studied, and Planck's constant and work function of different electrode materials are analyzed by using photoelectric effect.

## THEORY AND PHYSICAL PRINCIPLES

Surface electrons in a metal can be emitted out by exposing the metal surface to monochromatic electromagnetic (EM) radiation. The emitted electrons are called photoelectrons. This process has several important features that are not consistent with classical physics; a) there is no lag time, which means electrons can emit immediately after exposure, b) kinetic energy of the emitted electrons are independent of intensity of EM radiation, c) photoelectron emission stats after certain frequency of EM radiation, which is called cut off frequency.

In 1905 Albert Einstein postulated a relationship of energy of the EM radiation ($E_f$), kinetic energy of photoelectrons (K) and effect of cut off frequency as a work-function of metal ($\phi$).

$$E_f = K + \phi \tag{1}$$

This postulate also confirmed Planck's hypothesis of energy quanta in electromagnetic radiation. In which EM radiation carries energy as a quanta or packets.

$$E_f = hf \tag{2}$$

$f$ is frequency of the EM radiation and $h$ is Planck's constant which is the most fundamental constant in quantum physics.

$$h = 6.626 \times 10^{-34} \, J \cdot s = 4.136 \times 10^{-15} \, eV \cdot s \tag{3}$$

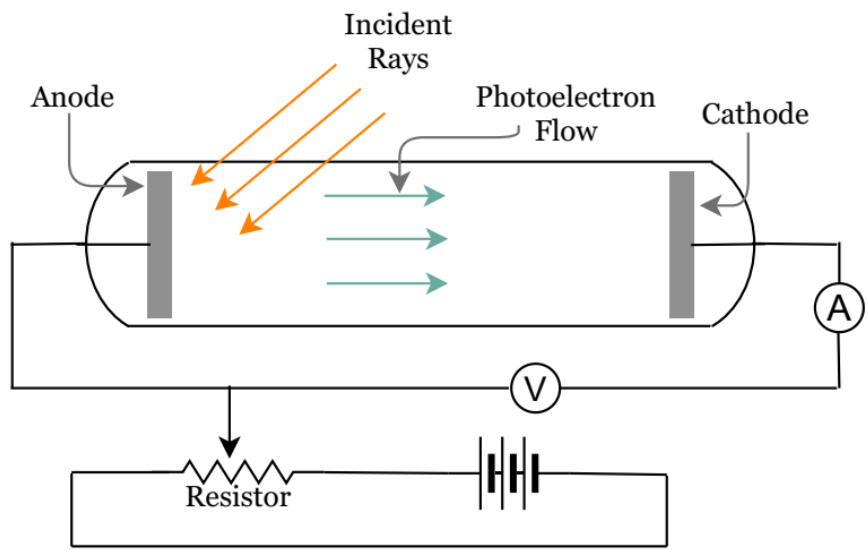

Figure 1    Schematic diagram of photoelectric effect setup

Kinetic energy of the photoelectrons,

$$K = hf \ - \ \phi \tag{4}$$

Einstein elaborate physics of equation (4) very in detailed to explain the meaning of work function ($\phi$). Kinetic energy of photoelectrons can have only positive values, which means there should be a minimum frequency of photons (EM radiation) to create photoelectrons.

Threshold frequency or cut off frequency when the kinetic energy of photoelectrons with zero kinetic energy,

$$f_c = \frac{\phi}{h} \tag{5}$$

Cut off wavelength of EM wave,

$$\lambda_c = \frac{c}{f_c} = \frac{c}{\phi/h} = \frac{hc}{\phi} \tag{6}$$

Speed of light, $c = 2.99 \times 10^8 \ \frac{m}{s}$ and $hc \ = \ 1240 \ eV \cdot nm$

Work function of the material can be found by applying a stopping potential ($\Delta V$) from an external circuit as shown in figure 1.

$$K = U_{electric} = \Delta V = \frac{h}{e}f \ - \ \frac{\phi}{e} \tag{7}$$

Slope of the graph of stopping potential vs frequency can be used to find the Planck's constant ($h$) and the intercept of the graph can be used to find the work function ($\phi$) of the electrode material.

**APPARATUS AND PROCEDURE**

- This experiment is done with the following simulation: https://phet.colorado.edu/sims/cheerpj/photoelectric/latest/photoelectric.html?simulation=photoelectric
- A very detail lesson video with detail of data collection with simulator and data analysis with excel can be found here: https://youtu.be/6blKJc36scw

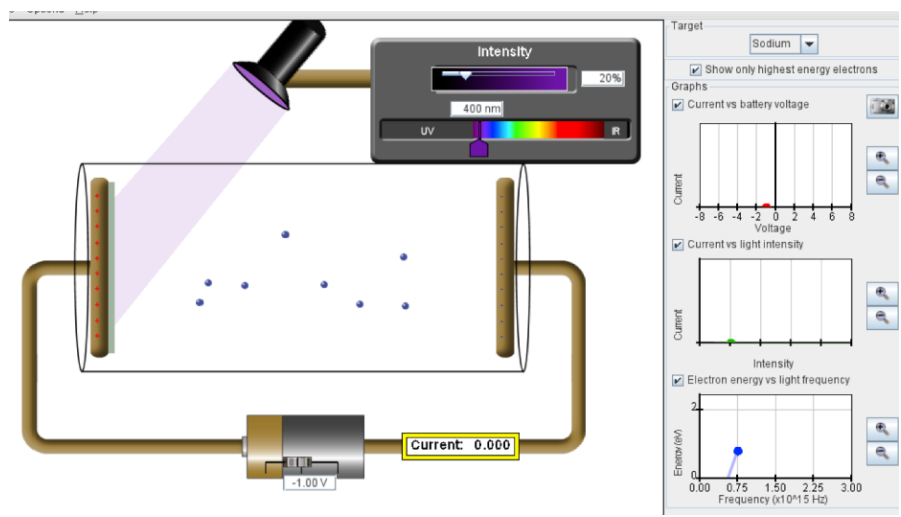

Figure 2    Simulation for photoelectric effect experiment (Picture credit: https://phet.colorado.edu/)

*Part A: Investigation of quantum model of EM radiation*

- This can be tested by checking the effect of the intensity of incident EM radiation.
- Set the frequency of the EM wave to high-UV and do not change the frequency of the EM wave during this part of the experiment.
- Set the target (by using a selection tool on top right side) into material of electrode given in table 1.
- Make sure that the external battery voltage creates the positive end at the electrode which exposes to EM radiation.
- Slowly increase the stopping potential from the external voltage supply (battery) till the photoelectron stops just before the negative electrode and bounces back to the starting electrode. This is the exact point when the photocurrent is zero.
- Note down the stopping potential by increasing intensity of EM radiation.
- Change the electrode material and repeat the above procedure.

*Part B: Investigation the Planck's constant by using Photoelectric effect*

- Set the intensity of the incident radiation to fixed value (100%). Technically it does not matter but keep it fixed during the rest of the experiment. Technically it does not have any effect on the experiment.
- Set the electrode material to Sodium (first one in the selection tool).
- Measure the stopping potential for highest UV radiation. This is the point when the photocurrent becomes zero.
- Then reduce the wavelength (about 50.0 nm at a time) of the radiation and then find the stopping potential.
- Then repeat the above procedure for different electrode materials.
- Make a graph of stopping potential vs frequency of EM radiation.
- Fit the data with linear fitting and use the slope of the graph to find the Planck's constant and intercept of the graph to find the work function of the electrode material.
- Discuss the calculated Planck's constant and work-function in terms of quantum model?
- Change the electrode material and repeat the above procedure.

**PRE LAB QUESTIONS**

1) Describe the concept of energy quanta of EM radiation which was explained by Planck?
2) Describe work function in photoelectric effect?
3) Describe cut off frequency in photoelectric effect?
4) Describe how to identify unknown EM radiation by using photoelectric effect?
5) Describe the solar cell in terms of application of photoelectric effect?

## DATA ANALYSIS AND CALCULATIONS

*Part A: Investigation of quantum model of EM radiation*

Table 1    Effect of intensity or number of photons of incident EM radiation

| Percentage intensity or number of photons of incident EM radiation Use mid UV for this study | Stopping potential for Sodium [    ] Wavelength of EM wave λ = [    ] | Stopping potential for Zinc [    ] Wavelength of EM wave λ = [    ] | Stopping potential for Calcium [    ] Wavelength of EM wave λ = [    ] | Stopping potential for unknown [    ] Wavelength of EM wave λ = [    ] |
|---|---|---|---|---|
| 20 | | | | |
| 50 | | | | |
| 70 | | | | |
| 100 | | | | |

*Part B: Investigation the Planck's constant by using Photoelectric effect*

o   Make a graph of stopping potential vs frequency of EM radiation.
o   Fit the data with linear fitting.
o   Use the slope of the graph to find the Planck's constant.
o   Use the intercept to find the work function of the electrode material.
o   Discuss the calculated Planck's constant and work-function in terms of quantum model?

Table 2      Stopping potential for each color and analysis of work function of electrode

| Electrode material 1: Sodium | | | Electrode material 2: Zinc | | |
|---|---|---|---|---|---|
| Wavelength [   ] | Frequency [     ] | Stopping Potential [     ] | Wavelength [   ] | Frequency [     ] | Stopping Potential [     ] |
| | | | | | |
| | | | | | |
| | | | | | |
| | | | | | |
| | | | | | |
| Electrode material 3: Calcium | | | Electrode material 4: Unknown | | |
| Wavelength [   ] | Frequency [   ] | Stopping Potential [     ] | Wavelength [   ] | Frequency [     ] | Stopping Potential [     ] |
| | | | | | |
| | | | | | |
| | | | | | |
| | | | | | |
| | | | | | |

Table 3   Analysis of Planck's constant and work function

| Electrode material | Slope of the graph $h/e$ [     ] | y-intercept of graph, $\frac{\phi}{e}$ [     ] | Work function of material, $\phi$ [     ] | Planck constant calculated, $h_{cal}$ [     ] | PE between $h_{cal}$ and $h_{known}$ [     ] |
|---|---|---|---|---|---|
| Sodium | | | | | |
| Zinc | | | | | |
| Calcium | | | | | |
| Unknown | | | | | |

# EXPERIMENT 11    ATOMIC SPECTRA AND BOHR MODEL

## OBJECTIVE

Spectral lines of emission spectra are studied. Rydberg formula is investigated by using the Hydrogen emission spectrum. Emission spectrums of other elements are studied.

## THEORY AND PHYSICAL PRINCIPLES

One of the early developments of quantum physics was started with the observation of spectrums. There are two types of spectrums, a) absorption spectrum and b) emission spectrum. Figure 1(a) shows an experimental setup to observe absorption spectrum, in which light passes through a gas tube and some energy is absorbed by the gras and those absent energy can be observed in the back lines in the absorption spectrum. Figure 1(b) shows an experimental setup to observe emission spectrum, in which light emitted from a gas discharge tube generates a spectrum. Both spectrums consist of intrinsic information of the electronic structure of the gas medium.

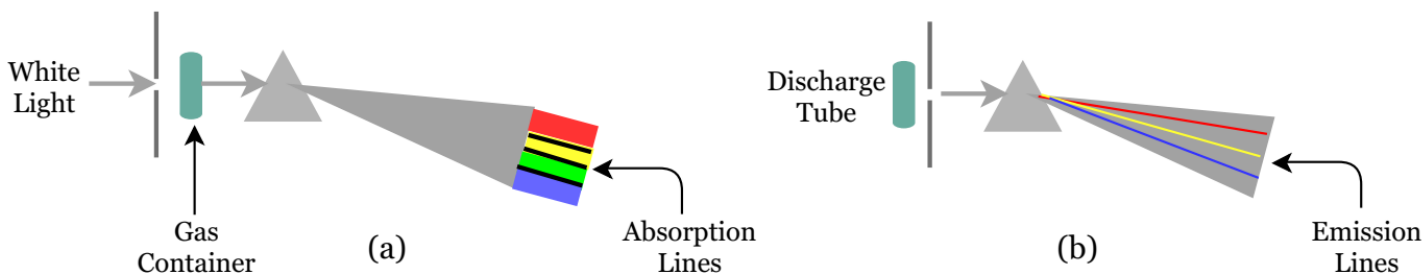

Figure 1   Schematic diagrams of experimental setup to observe (a) absorption spectrum, and (b) emission spectrum

In 1885 Johann Balmer proposed an empirical formula describe the wavelength ($\lambda$) of the observed color of the hydrogen emission spectrum and it is called the Balmer series for Hydrogen.

$$\frac{1}{\lambda} = R_H \left( \frac{1}{2^2} - \frac{1}{n^2} \right) \tag{1}$$

$R_H = 1.09737 \times 10^7 \ m^{-1}$ is called Rydberg constant for Hydrogen and $n$ = 3, 4, 5, 6, ............

Other series of emission lines were detected on the Hydrogen spectrum in the 20th century and all of them were successfully explained by Rydberg formula.

$$\frac{1}{\lambda} = R_H \left( \frac{1}{n_f^2} - \frac{1}{n_i^2} \right) \tag{2}$$

$n_i$ = $n_f$+1, $n_f$+2, $n_f$+3, ....................,  and $n_f$ is any positive integer
$n_f$ = 1; Lyman series, $n_f$ = 2; Balmer series, $n_f$ = 3; Paschen series, $n_f$ = 4; Brackett series, $n_f$ = 5; Pfund series, $n_f$ = 6; Humphreys series

The Rydberg formula was a mystery until it was explained theoretically by the Bohr model in 1913. Rydberg constant was successfully calculated by Bohr model, which is developed by using three postulates as follows.

1. Negative electrons move around the positive nucleus and all electron orbits are circles.

2. Electron angular momentum must satisfy the first quantization condition.

$$L_n = n\hbar, \text{ where } n = 1, 2, 3, \ldots\ldots \tag{3}$$

$$\hbar = \frac{h}{2\pi}, \quad h = 6.626 \times 10^{-34} \, J \cdot s = 4.136 \times 10^{-15} \, eV \cdot s \tag{4}$$

By using classical physics circular motion of electron,

$$L_n = n\hbar = m_e v_n r_n \tag{5}$$

$m_e$ is mass of electron, $m_e = 9.11 \times 10^{-31} \, kg$, $v_n$ is orbital speed of electron in $n^{th}$ orbit, $r_n$ is orbital radius of electron in nth orbit.

3. Electrons are allowed to transfer from $n^{th}$ orbit $m^{th}$ orbit. When atoms absorb energy electrons can transfer to higher energy states and when an excited electron transit back to lower energy states then energy is emitted. This says absorb or emitted energy must be quantized.

$$\Delta E = |E_n - E_m| = hf \tag{6}$$

$hf$ is the energy of the absorb/emitted photon and the $f$ is the frequency of the photon.

These postulates combine with classical physics and circular motion and concept of energy used to develop the first model of hydrogen atom.

Radius of electron orbit, $r_n = a_0 n^2$ \hfill (7)

$$a_0 = \frac{4\pi\varepsilon_0\,\hbar^2}{m_e e^2} = 5.29 \times 10^{-11} \, m = 0.529 \, \text{Å} \tag{8}$$

Mass of electron, $e = 1.602 \times 10^{-19} \, C$ and permittivity constant, $\varepsilon_0 = 8.85 \times 10^{-12} \, \frac{C^2}{Nm^2}$

Orbital energy of an electron, $E_n = -E_0 \frac{1}{n^2}$ \hfill (9)

$$E_0 = \frac{1}{32\pi^2\varepsilon_0^2}\frac{m_e e^2}{\hbar^2} = 2.17 \times 10^{-18} \, J = 13.6 \, eV \tag{10}$$

Spectral emission lines of Hydrogen due to photons emitted due to electron transfer from excited state to lower energy state.

$$\Delta E = hf = |E_n - E_m| = E_0 \left(\frac{1}{m^2} - \frac{1}{n^2}\right) \tag{11}$$

$$\frac{1}{\lambda} = \frac{E_0}{hc}\left(\frac{1}{m^2} - \frac{1}{n^2}\right) \tag{12}$$

$$R_H = \frac{E_0}{hc} = 1.097 \times 10^7 \, \frac{1}{m} \tag{13}$$

## APPARATUS AND PROCEDURE

- This experiment is done with following simulation:
  https://phet.colorado.edu/sims/cheerpj/discharge-lamps/latest/discharge-lamps.html?simulation=discharge-lamps

- A very detail lesson video with detail of data collection with simulator and data analysis with excel can be found here: https://youtu.be/C5Jh3bHgPfo

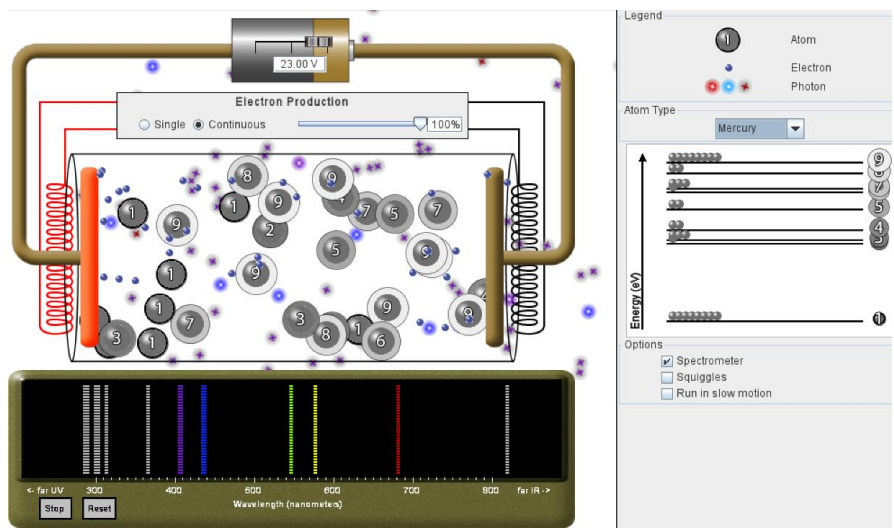

Figure 2    Simulation to study emission spectrum (Picture credit: https://phet.colorado.edu/)

- Set the battery voltage to maximum (+30.0 V).
- Set the electron production to "continues".
- Set the atom type to Hydrogen in the selection panel top right side of the selection tool.
- Click "spectrometer" and "run in slow motion" on the bottom right side of the selection panel.
- Spectrometer shows all the emission spectral lines of Hydrogen atoms.
- When an electron collides with the atom, excitation can be observed on the energy state diagram on the right corner of the simulator.
- Observe carefully the excited state number (initial state = m) and then when the deexcitation happens observe carefully the next state (final state = n) and the color of the emitted photon.
- Note down the initial and final states (m and n) and the wavelength of observed color from the spectrum tool.
- Repeat above procedure for all the other observable colors of the hydrogen spectrum.
- Then repeat all the above procedures for other types of elements by changing the "atom type".

### PRE LAB QUESTIONS

1) Describe absorption and emission spectra?
2) Describe the Rydberg equation?
3) Describe Hydrogen model?

**DATA ANALYSIS AND CALCULATIONS**

Table 1   Atomic spectra Hydrogen spectrum

| Emitted EM radiation | Wavelength $\lambda$ [     ] | Excited state m | De-excited state n | $\frac{1}{m^2} - \frac{1}{n^2}$ | Wavelength $1/\lambda$ [     ] |
|---|---|---|---|---|---|
|  |  |  |  |  |  |
|  |  |  |  |  |  |
|  |  |  |  |  |  |
|  |  |  |  |  |  |

Table 2   Analysis of Rydberg and Planck's constants

| Emitted EM radiation | Rydberg Constant $R_H$ [     ] | PE of Rydberg constant | Plank's Constant $R_H$ [     ] | PE of Planck's constant |
|---|---|---|---|---|
|  |  |  |  |  |
|  |  |  |  |  |
|  |  |  |  |  |
|  |  |  |  |  |

- Find the average of the Rydberg constant ($R_{cal\_avg}$) calculated in table 1?
- Compare (find PE) calculated average and the known values of Rydberg constants?
- Make a graph of $1/\lambda$ vs $\left(\frac{1}{m^2} - \frac{1}{n^2}\right)$?
- Find the Rydberg constant ($R_{graph}$) by using slope of the graph?
- Compare (find PE) $R_{graph}$ and $R_{known}$?
- Compare (find PE) $R_{graph}$ and $R_{cal\_avg}$?
- Find the Planck's constant by using the calculated average Rydberg constant and compare it with the known value of Planck's constant?
- Find the Planck's constant by using the Rydberg constant from the graph and compare it with the known value of Planck's constant?

Table 3     Percent error analysis of Rydberg and Planck's constants

|  | Rydberg constant [   ] | PE [   ] | Planck's constant [   ] | PE [   ] |
|---|---|---|---|---|
| Calculated average |  |  |  |  |
| From graph |  |  |  |  |

Table 4     Atomic spectra of Mercury

| Emitted EM radiation | Wavelength λ [     ] | Excited state m | De-excited state n | Check the validity Rydberg formula $\dfrac{1}{\lambda\left(\dfrac{1}{m^2}-\dfrac{1}{n^2}\right)}$ [     ] |
|---|---|---|---|---|
|  |  |  |  |  |
|  |  |  |  |  |
|  |  |  |  |  |
|  |  |  |  |  |
|  |  |  |  |  |

Table 5     Atomic spectra of Sodium

| Emitted EM radiation | Wavelength λ [     ] | Excited state m | De-excited state n | Check the validity Rydberg formula $\dfrac{1}{\lambda\left(\dfrac{1}{m^2}-\dfrac{1}{n^2}\right)}$ [     ] |
|---|---|---|---|---|
|  |  |  |  |  |
|  |  |  |  |  |

- Discuss the validity of Rydberg formula for heavy elements?

# EXPERIMENT 12     RADIATION AND RADIOACTIVE HALF-LIFE

## OBJECTIVE

Radiation of radioactive isotope is studied. Radiation shielding of different types of radiation is investigated in terms of type of material and thickness of material. Radioactive half-life is calculated.

## THEORY AND PHYSICAL PRINCIPLES

Some of the elements in the periodic table are not stable and they tend to decay into more stable nuclei over time. This process is called radioactive decay. Radioactive decaying process of an element depends on the number of decaying processes per time, which is also known as count rate.  When the decaying of one nucleus happens in the material it emits a particle, and that particle can be measured by using a Geiger counter.

$$\frac{\Delta N}{\Delta t} \propto N \tag{1}$$

By considering the instantaneous rate of change,

$$\frac{dN}{dt} = -\lambda N \tag{2}$$

$\lambda$ is decay constant for the radioactive isotope.

$$\int_{N_0}^{N} \frac{dN}{N} = -\lambda \int_{0}^{t} dt \tag{3}$$

$N_0$ is the amount when the time $t$=0 and by integrating above equation.

$$N(t) = N_0 e^{-\lambda t} \tag{4}$$

Half-life of the radioactive isotope is the time it takes to change the starting amount of detected counts by half. When, N($t$)=N$_0$/2, $t$ = 1 half-life = $\tau$

$$N(t = \tau) = \frac{N_0}{2} = N_0 e^{-\lambda \tau} \tag{5}$$

$$e^{\lambda \tau} = 2 \tag{6}$$

$$\tau = \frac{ln2}{\lambda} = \frac{0.693}{\lambda} \tag{7}$$

To find the half-life ($\tau$) of a radioactive isotope it is required to find the decay constant ($\lambda$). It can be done by measuring radioactivity of an isotope as a function of time.

$$e^{-\lambda t} = \frac{N}{N_0} \tag{8}$$

$$ln\,N = -\lambda\,t + lnN_0 \tag{9}$$

Slope of the graph of $ln\,N$ vs time is the decay constant and intercept the graph can be used to find out the starting radiation count of material when the count started at time equal to zero. Usually this lab is done with an apparatus called a Geiger counter, which can detect the emission of radioactive material.

**APPARATUS AND PROCEDURE**

- This experiment is done with following simulation: https://www.gigaphysics.com/gmtube_lab.html
- A very detail lesson video with detail of data collection with simulator and data analysis with excel can be found here: https://youtu.be/rpNBPwo2YSw

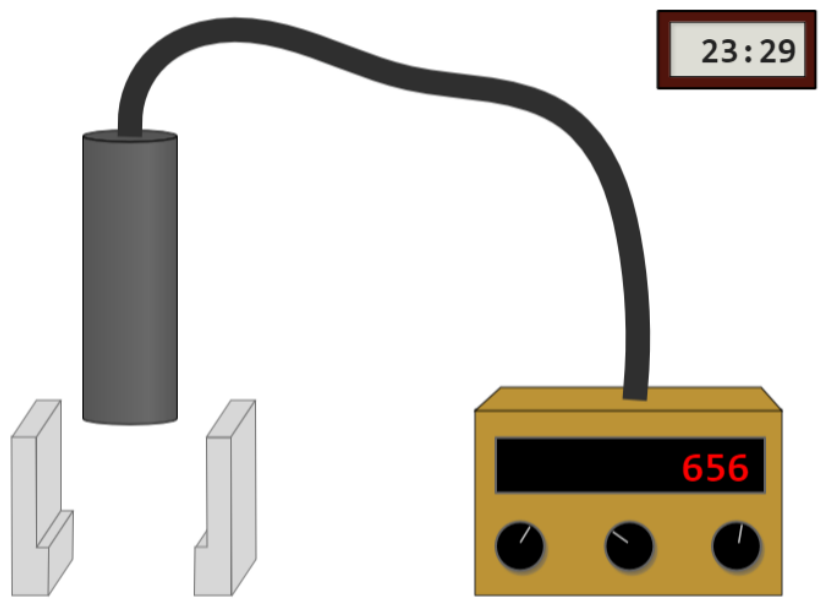

Figure 1  Simulation for the radiation and radioactive half-life experiment
(Picture credit: https://www.gigaphysics.com/)

*Part A: Study types of radiation and radiation shielding*

- Set the radiation source to "background". Set the type of barrier to "cardboard". Set the number of barriers to "one". Set the time duration to 30 seconds.
- Click on "start count" and note down the total count per 30 seconds.
- Increase the number of barriers by one at a time and repeat the above procedure.
- Then change the barrier material to plastic and repeat the radiation count with an increasing number of barriers.
- Then change the barrier material to lead and repeat the radiation count with increasing number of barriers.
- Change the radiation source to "alpha". Set the type of barrier to "cardboard". Set the number of barriers to "one". Set the time duration to 30 seconds.
- Click on "start count" and note down the total count per 30 seconds.
- Increase the number of barriers by one at a time and repeat the above procedure.
- Then change the barrier material to plastic and repeat the radiation count with an increasing number of barriers.
- Then change the barrier material to lead and repeat the radiation count with increasing number of barriers.
- Change the radiation source (beta, gamma and unknown) and repeat the procedure above.
- Discuss radiation shielding with respect to type and thickness of barrier material for each of the radiation types above.
- Identify the unknown radiation source by comparing the results with the observation of known radiation sources.

*Part B: Study of half-life of radioactive materials*

- Change the radiation source to "Ba-137". Set the type of barrier to "none". Set the number of barriers to "none". Set the time duration to 30 seconds.
- Click on "New Ba-137 source" on the right side of the selection panel.
- Click on "start count" and note down the total count after 30 seconds.
- Repeat the last step above for 10 more times.
- Find the total count as the time increases.
- Find the natural log of each 30 seconds (ln N).
- Make a graph of counts per 30 seconds vs time.
- Discuss the behavior of the graph in terms of radioactive decay.
- Make a graph of *ln*N vs time and fit the graph with linear fitting.
- Find the radioactive decay constant by using slope of the graph and the starting count when *t*=0 by using intercept of the graph.
- Find the radioactive half-life of Ba-137 by using decay constant.

**PRE LAB QUESTIONS**

1) Describe the radioactive decay?
2) Describe radioactive half-life?
3) Describe background radiation?
4) Describe alpha radiation?
5) Describe beta radiation?
6) Describe gamma radiation?
7) Describe nuclear fusion?
8) Describe nuclear fission?

## DATA ANALYSIS AND CALCULATIONS

*Part A: Study of types of radiation and radiation shielding*

Table 1   Radiation rate of different types of radiation and shielding effects

| Radiation source: Background | | | | | |
| --- | --- | --- | --- | --- | --- |
| Number of Cardboards | Count per 30 seconds | Number of Plastic | Count per 30 seconds | Number of Lead | Count per 30 seconds |
|  |  |  |  |  |  |
|  |  |  |  |  |  |
|  |  |  |  |  |  |

| Radiation source: Alpha | | | | | |
| --- | --- | --- | --- | --- | --- |
| Number of Cardboards | Count per 30 seconds | Number of Plastic | Count per 30 seconds | Number of Lead | Count per 30 seconds |
|  |  |  |  |  |  |
|  |  |  |  |  |  |
|  |  |  |  |  |  |

| Radiation source: Beta | | | | | |
| --- | --- | --- | --- | --- | --- |
| Number of Cardboards | Count per 30 seconds | Number of Plastic | Count per 30 seconds | Number of Lead | Count per 30 seconds |
|  |  |  |  |  |  |
|  |  |  |  |  |  |
|  |  |  |  |  |  |

| Radiation source: Gamma | | | | | |
| --- | --- | --- | --- | --- | --- |
| Number of Cardboards | Count per 30 seconds | Number of Plastic | Count per 30 seconds | Number of Lead | Count per 30 seconds |
|  |  |  |  |  |  |
|  |  |  |  |  |  |
|  |  |  |  |  |  |

| Radiation source: Unknown | | | | | |
| --- | --- | --- | --- | --- | --- |
| Number of Cardboards | Count per 30 seconds | Number of Plastic | Count per 30 seconds | Number of Lead | Count per 30 seconds |
|  |  |  |  |  |  |
|  |  |  |  |  |  |
|  |  |  |  |  |  |

- Discuss radiation shielding with respect to type and thickness of barrier material for each of the radiation types above.
- Identify the unknown radiation source by comparing the results with known radiation sources.

*Part B: Study of half-life of radioactive materials*

Table 2    Radioactive half-life of Ba-137 isotope

| Time, t<br>[     ] | Count per minute<br>[     ] | Total count<br>[     ] | *ln* N<br>[     ] |
|---|---|---|---|
|  |  |  |  |
|  |  |  |  |
|  |  |  |  |
|  |  |  |  |
|  |  |  |  |
|  |  |  |  |
|  |  |  |  |
|  |  |  |  |
|  |  |  |  |
|  |  |  |  |

- Make a graph of count per 30 seconds vs time.
- Discuss the behavior of the graph in terms of radioactive decay.
- Make a graph of *ln*N vs time and fit the graph with linear fitting.
- Find the radioactive decay constant by using slope of the graph and the starting count when *t*=0 by using intercept of the graph.
- Find the radioactive half-life of Ba-137 by using decay constant.